\documentclass[10pt,journal,compsoc]{IEEEtran}

\usepackage[T1]{fontenc}

\ifCLASSINFOpdf
\else
\fi

\usepackage{xcolor,colortbl}
\usepackage{float}
\usepackage{graphicx}
\usepackage{subfigure}
\usepackage{multirow}
\usepackage{tcolorbox}
\usepackage{xcolor}
\usepackage{colortbl}
\usepackage{diagbox}
\usepackage{bm}

\usepackage{amssymb}

\usepackage{pifont}
\usepackage{setspace}
\usepackage{color}
\usepackage{algpseudocode}
\usepackage{hyperref}
\usepackage{framed}
\usepackage{amsmath}
\usepackage{enumitem}
\usepackage{balance}
\usepackage{url}
\usepackage[T1]{fontenc}
\usepackage[utf8]{inputenc}
\usepackage[font=small,labelfont=bf,tableposition=top]{caption}
\usepackage{booktabs}
\usepackage{threeparttable}
\usepackage[ruled,vlined,linesnumbered]{algorithm2e}

\usepackage[noadjust]{cite}

\definecolor{grey}{rgb}{0.9,0.9,0.9}
\definecolor{lightgreen}{HTML}{bae4b3}
\definecolor{lightgrey}{HTML}{f0f0f0}
\definecolor{mygreen}{HTML}{31a354}
\definecolor{mygray}{HTML}{666666}

\hypersetup{
	colorlinks=true,      
	linkcolor=black,       
	citecolor=black,    
	filecolor=cyan,       
	urlcolor=black          
}

\makeatletter

\newcommand{\Rmnum}[1]{\expandafter\@slowromancap\romannumeral #1@}
\makeatother

\newcommand*{\mycode}{\fontfamily{lmtt}\selectfont}

\begin{document}
%
\title{Runtime Permission Issues in Android Apps: Taxonomy, Practices, and Ways Forward}
%
%
%

\author{Ying Wang, Yibo Wang, Sinan Wang, Yepang Liu\IEEEauthorrefmark{2}, Chang Xu~\IEEEmembership{Senior Member,~IEEE}, \\Shing-Chi Cheung~\IEEEmembership{Senior Member,~IEEE}, Hai Yu, and Zhiliang Zhu

\thanks{Ying Wang, Yibo Wang, Hai Yu, and Zhiliang Zhu are with the Software College, Northeasthern University, China. E-mail: wangying@swc.neu.edu.cn, wangc\_neu@163.com, \{yuhai, zzl\}@mail.neu.edu.cn.}
\thanks{Sinan Wang and Yepang Liu are with the Department of Computer Science and Engineering, and Guangdong Provincial Key Laboratory of Brain-inspired Intelligent Computation, Southern University of Science and Technology, Shenzhen, China. E-mail: wangsn@mail.sustech.edu.cn, liuyp1@sustech.edu.cn.}
\thanks{Chang Xu is with the State Key Lab for Novel Software Technology and the Department of Computer Science and Technology, Nanjing University, China. E-mail: changxu@nju.edu.cn.}
\thanks{Shing-Chi Cheung is with the Department of Computer Science and Engineering, The Hong Kong University of Science and Technology, Hong Kong, China. E-mail: scc@cse.ust.hk.}
\thanks{\IEEEauthorrefmark{2} Yepang Liu is the corresponding author.}
\thanks{Manuscript received 2021.}
}

\IEEEtitleabstractindextext{%
	\begin{abstract}
		Android introduces a new permission model that allows apps to request permissions at runtime rather than at the installation time since 6.0 (Marshmallow, API level 23).
		While this runtime permission model provides users with greater flexibility in controlling an app's access to sensitive data and system features, it brings new challenges to app development. 
		First, as users may grant or revoke permissions at any time while they are using an app, developers need to ensure that the app properly checks and requests required permissions before invoking any permission-protected APIs.
		Second, Android's permission mechanism keeps evolving and getting customized by device manufacturers.
		Developers are expected to comprehensively test their apps on different Android versions and device models to make sure permissions are properly requested in all situations.
		Unfortunately, these requirements are often impractical for developers. In practice, many Android apps suffer from various runtime permission issues (ARP issues). 
		
		While existing studies have explored ARP issues, the understanding of such issues is still preliminary.
		To better characterize ARP issues, we performed an empirical study using 135 Stack Overflow posts that discuss ARP issues and 199 real ARP issues archived in popular open-source Android projects on GitHub.
		Via analyzing the data, we observed 11 types of ARP issues that commonly occur in Android apps.
		For each type of issues, we systematically studied: (1) how they can be manifested, (2) how pervasive and serious they are in real-world apps, and (3) how they can be fixed. 
		We also analyzed the evolution trend of different types of issues from 2015 to 2020 to understand their impact on the Android ecosystem.
		Furthermore, we conducted a field survey and in-depth interviews among practitioners, to gain insights from industrial practices and learn practitioners' requirements of tools that can help combat ARP issues.
		Finally, to understand the strengths and weaknesses of the existing ARP issue detectors, we built \textsc{ARPBench}, an open benchmark consisting of 94 real ARP issues, and evaluated the performance of three available ARP issue detectors.
		The experimental results indicate that the existing tools have very limited supports for detecting our observed issue types and report a large number of false alarms.
		We further analyzed the tools' limitations and summarized the challenges of designing an effective ARP issue detection technique.
		We hope that our findings can shed light on future research and provide useful guidance to practitioners.
		
	\end{abstract}
	
	
}

\maketitle


%
\IEEEpeerreviewmaketitle


\section{Introduction}
\label{sec:Introduction}

\IEEEPARstart{T}{he} permission mechanism plays a significant role in securing Android apps~\cite{felt2011android}.
According to the official Android documentation~\cite{Android_document},
to access sensitive user data (e.g., photos and contacts) or critical system features (e.g., camera and GPS), an Android app needs to request the corresponding \textit{dangerous permissions} first.
Prior to Android 6.0 (Marshmallow, API level 23)
, users must grant all permissions that an app asks for at the installation time in order to successfully install the app~\cite{andriotis2016permissions, feng2019ac}.
Once installed, the app will be able to access all permitted user data or system features without further permission checks~\cite{alepis2017hey}.
This ``static'' permission model was vulnerable and caused many undesirable consequences~\cite{bao2017automated}.
For security enhancement, Android~6.0 introduced a \textit{runtime permission model} that lets apps request permissions at runtime instead of at the installation time.
With the new model, users can approve or deny an app's permission requests at their discretion.
Additionally, users are allowed to grant or revoke permissions anytime via changing system settings.

While the runtime permission model improves system security, permission handling has become a non-trivial task for app developers since Android~6.0, mainly due to the following three reasons:
\begin{itemize}
	\item \textbf{Unpredictable user interactions:} First of all, users' permission-control actions can be highly unpredictable.
	If users deny or revoke the permissions required by certain APIs, apps using such \textit{permission-protected APIs} may easily crash at runtime.
	Thus, app developers should anticipate all possible user actions and carefully check whether the required permissions have been granted whenever the permission-protected APIs are invoked to avoid buggy app behaviors.
	This is not an easy task for developers.
	For example, in post 34150083 ~\cite{post34150083} on Stack Overflow~\cite{SO} (SO for short), developers discussed when to request runtime permissions to avoid crashes.
	One accepted answer is that ``\textit{In general, request needed permissions as soon as you need them. This way you can inform the user why you need the permission and handle permission denies much easier.
		Think of scenarios where the user revokes the permission while your app runs: If you request it at startup and never check it later this could lead to unexpected behaviour or exceptions.}''
	
	\item \textbf{Frequent evolution of APIs and permissions:} Second, the Android framework keeps evolving, which often introduces or removes permission-protected APIs and dangerous permissions.
	Developers should continually adapt their app to avoid unexpected issues when running on new Android versions.
	This is also a challenging task for developers.
	If they do not actively track the evolution of APIs and permissions, they may often make mistakes.
	Take the issue discussed in SO post 58428763~\cite{post58428763} as an example.
	The developer found that the Bluetooth scanning function of his/her app works on all devices except those with Android 10.
	The answer with the highest up votes revealed the root cause of the issue: Bluetooth scanning function requires an additional permission {\mycode ACCESS\_BACKGROUND\_LOCATION} since Android 10.0.
	This post has been viewed more than 18k times, suggesting that many
	developers may have encountered similar issues.

	\item \textbf{Customized permission mechanism:} Third, Android device manufacturers (e.g., Samsung and Huawei) may customize the permission mechanism of the vanilla Android when building their own mobile devices.
	Such customization can cause confusion to developers.
	For instance, in SO post 52781910~\cite{post52781910}, a developer asked how to deal with the permission ``\emph{show on lock screen}'', which is customized on \emph{Xiaomi} devices. 
	As we will show later, such issues induced by manufacturers' customization
	are common, causing immense frustration to app developers.

\end{itemize}

Because of the above-mentioned challenges, developers can easily make mistakes
in permission handling. These mistakes result in various issues in Android
apps. We refer to such issues stemming from the improper handling of runtime
permissions as \textit{Android runtime permission issues} (ARP issues for
short).
Existing studies~\cite{dilhara2018automated, bogdanas2017dperm,
	fang2016revdroid} have explored ARP issues, but focused only on specific issue
types. The issues are explored to ease app migration from the static permission
model prior to Android 6.0 to the new runtime permission model.
However, as we will show later, such explored issues are only a tip of the
iceberg.
The wide landscape of Android versions and device models has given rise to
different types of ARP issues. 
Yet, there is no systematic study of ARP issues.
The lack of understanding of the issues would impede the development of useful techniques to help developers deal with various ARP issues.
This motivates us to conduct a systematic research on ARP issues via three
studies as shown in Table 1. 
We briefly introduce the three studies below.

\begin{table}[]
	\bgroup
	\caption{An overview of our three studies}
	\begin{tabular}{c|l|l}
		\toprule
		\textbf{Study} &
		\textbf{Research question(s)} & \textbf{Approach}           \\ \hline \hline
		\#1 & RQ1\textendash RQ3                    & Empirical study             \\ \hline
		\#2 & RQ4\textendash RQ5                    & Survey and interviews \\ \hline
		\#3 & RQ6                        & Experiment evaluation             \\ \bottomrule
	\end{tabular}
	\egroup
	\label{overview}
\end{table}

First, we performed an empirical study of ARP issues in real-world Android apps
to examine the following three research questions:
\begin{itemize}
	\item \textbf{RQ1 (Issue types and causes)}: \textit{What are the common types of ARP issues in Android apps? How do the issues manifest themselves? What are their root causes?}

	\item \textbf{RQ2 (Pervasiveness and seriousness)}: \textit{How pervasive and serious are ARP issues in real-world Android projects?}

	\item \textbf{RQ3 (Fixing strategies)}: \textit{How do Android developers fix ARP issues? Are there common patterns?}
	
\end{itemize}

To address RQ1\textendash RQ3, we collected 135 SO posts discussing ARP issues with high \textit{views} and \textit{up votes}, and 199 ARP issues from 415 open-source Android projects on GitHub~\cite{github}.
Analyzing the collected data, we observed 11 types of ARP issues and six
commonly-adopted fixing strategies.
Besides the issue taxonomy, we also made many other interesting observations by investigating each type of issues.
For example, in addition to the improper handling of runtime permissions by an app itself, ARP issues can also be caused by the interference of third-party libraries.
Section~\ref{sec:empirical} reports our empirical findings in detail.

The above empirical study was performed based on the data collected from the open-source community.
To cross-validate our findings gained from studying RQ1\textendash RQ3, and further understand the challenges in handling runtime permissions and dealing with ARP issues, we conducted a field survey and interviews among industry practitioners, with the aim to explore the following two research questions:

\begin{itemize}
	\item \textbf{RQ4 (Industrial practices)}: \textit{How do industry practitioners handle runtime permissions in their Android projects?}
	
	\item \textbf{RQ5 (Challenges faced by practitioners)}: \textit{What are
		the common challenges faced by the Android practitioners in dealing with
		ARP issues?}
\end{itemize}

The survey and interviews provide a new perspective of ARP issues,
complementary to the empirical study. We will discuss the insights that we have
learned from the participants' real experiences and lessons in
Section~\ref{sec:Industry}.

Finally, in recent years, there have been several tools that can help detect ARP issues in Android apps.
With the issue taxonomy derived in our empirical study, we are intrigued to know how these tools would perform in practice.
To this end, we built \textsc{ARPBench}, an open benchmark consisting of 94 ARP issues collected from real-world projects and experimentally evaluated three ARP issue detectors, which are freely accessible to Android developers.
We investigate the following research question in the experiments.

\begin{itemize}
	\item \textbf{RQ6 (Performance of existing ARP issue detectors)} : \textit{How do existing ARP issue detectors perform in terms of issue detection capability and false alarm rate?}
\end{itemize}

Our experiment results show that none of the evaluated detectors support detecting all the 11 types of ARP issues found in our empirical study.
These detectors suffer from a high false alarm rate and miss many real issues, which would seriously hinder their adoption.
To help build better tools in the future, we provide a detailed analysis of the limitations of existing detectors using real-world ARP issues as examples in Section~\ref{sec:Implications for Future Research}.

In summary, our work makes four major contributions:

\begin{itemize}
	\item To the best of our knowledge, we conducted the first large-scale empirical study of ARP issues in Android apps.
	Based on our in-depth analysis of real issues from open-source projects, we built by far the most comprehensive taxonomy of ARP issues.
	Our empirical findings can shed light on future research and provide useful guidance to practitioners.
	
	\item We conducted a field survey and interviews to understand ARP issues from industry practitioners' perspective.
	The survey and interviews not only validated the findings of our empirical study, but also provided further insights that cannot be gained by analyzing the data from the open-source community.
	
	\item We evaluated three existing ARP issue detectors and provided a detailed analysis of their strengths and weaknesses to guide future tool development.
	
	\item To facilitate future research, we have released all our artifacts, including \textsc{ARPBench}, the first open benchmark for evaluating ARP issue detectors, on GitHub (http://arp-issues.github.io/).
	
\end{itemize}

\section{Background}
\label{sec:PRELIMINARIES}

This section presents the background of our work. We define four terms to ease discussions.
\begin{enumerate}
	\item \textit{Legacy platforms}: We refer to the Android versions prior to Marshmallow (6.0) as legacy platforms.
	\item  \textit{Legacy apps}: We refer to the Android apps that target legacy platforms as legacy apps.
	\item \textit{New platforms}: We refer to Android 6.0 and above as new platforms.
	\item \textit{New apps}: We refer to the Android apps that target new platforms as new apps.
\end{enumerate}

\subsection{Android Permission Mechanism}
\begin{figure}[t!]
	\centering
	\includegraphics[width=0.48\textwidth]{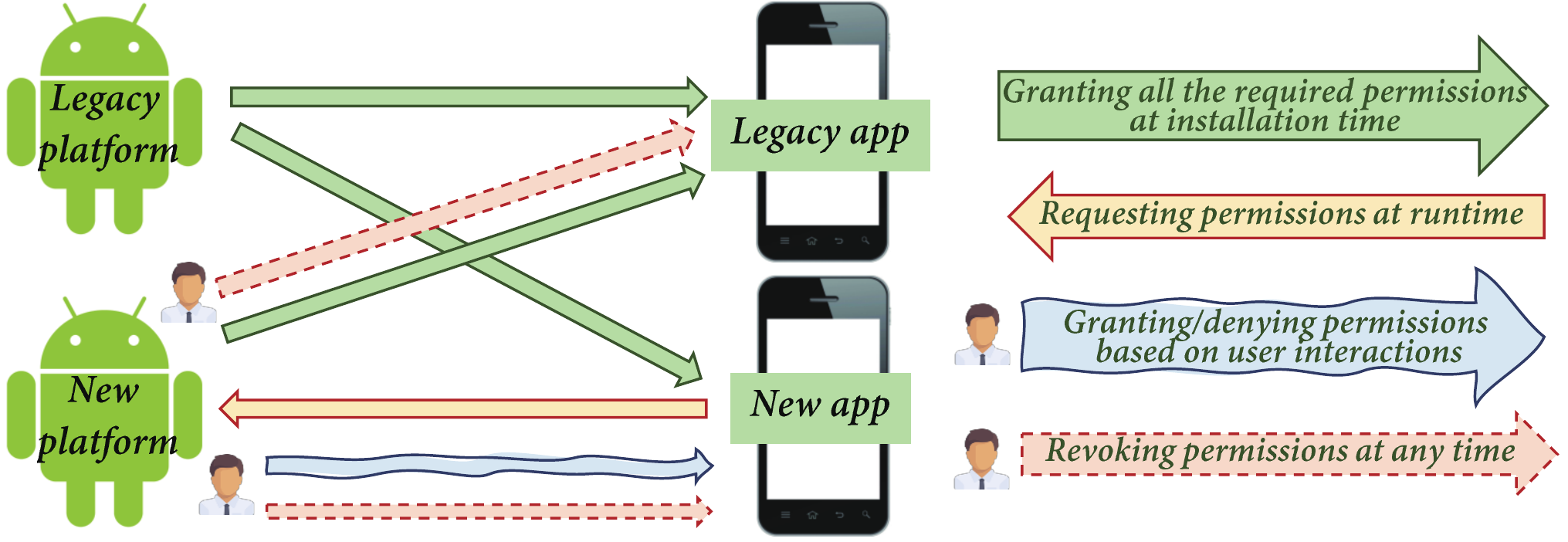}
	\caption{An overview of Android's permission mechanism}
	\label{background}
	\vspace{-4mm}
\end{figure}

As shown in Figure~\ref{background}, on legacy platforms, users are asked to grant the dangerous permissions required by legacy apps or new apps at the app installation time~\cite{alepis2017hey}.
Once the apps are installed, they will be able to access all permitted resources without further permission checks.
Under such a permission model, users lose control of permissions after installing apps and they have to uninstall the apps to gain the control back~\cite{zhauniarovich2016small}.
For safety concerns, Android switched to a runtime permission model starting from Marshmallow (Android 6.0).
Under this model, new apps need to check and request dangerous permissions at runtime by invoking designated APIs (see Section~\ref{sec:Request Process} for details) when they run on new platforms.
The new model also provides users with fine-grained control over the permissions: they can choose to approve or deny permission requests, and they can also revoke granted permissions at any time~\cite{dilhara2018automated}.
For backward compatibility, legacy apps running on new platforms still need to be granted all required permissions at installation.
However, users of new platforms are allowed to revoke the permissions granted to legacy apps. Such revocation could cause unexpected issues~\cite{bogdanas2017dperm} as we will discuss later.

\begin{figure*}[t!]
	\centering
	\includegraphics[width=0.999\textwidth]{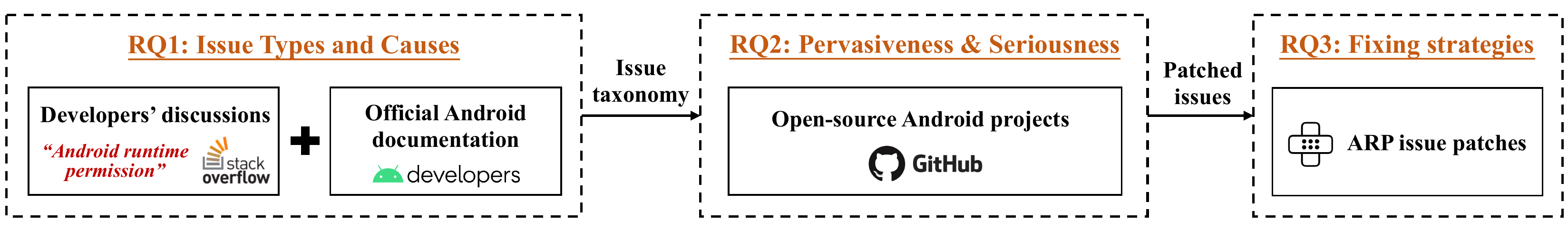}
	\caption{An overview of our empirical study on real-world ARP issues}
	\label{fig_overview}
\end{figure*} 

\subsection{Permission Check and Request Process}
\label{sec:Request Process}
Apps targeting new platforms should follow four steps to request dangerous permissions~\cite{Android_document}:

\textbf{\textit{Step 1: Declaration}.} They should first declare the required dangerous permissions in the {\mycode AndroidManifest.xml} file ({\mycode manifest} file for short), which is a configuration file to present essential information to build tools and the Android operating system.

\textbf{\textit{Step 2: Checking}.} Before invoking Android APIs that require dangerous permissions (i.e., \textit{permission-protected APIs}), the apps should check whether the corresponding permissions have already been granted by calling \textit{permission check APIs} provided by the Android platform. One example is {\mycode ContextCompat.checkSelfPermission()}. 
If the apps already have these permissions, they can proceed to invoke the permission-protected APIs.
Otherwise, they should request the permissions first.

\textbf{\textit{Step 3: Request}.} Starting from Marshmallow, the Android platform provides several \textit{permission request APIs} to allow apps to request dangerous permissions at runtime. One example is {\mycode ActivityCompat.requestPermissions()}.
Calling these APIs will bring up a dialog to interact with users, and users can choose to approve or deny the permission request.

\textbf{\textit{Step 4: Handling users' response}.} When the permission request dialog is closed after users make their choice, the callback method {\mycode onRequestPermissionsResult()} will be invoked by the Android platform to handle users' response. 
The apps need to properly override this method to find out whether the requested permissions have been successfully granted and take actions accordingly. 

According to the runtime permission model, all the above four steps are suggested to be performed by Android apps to request dangerous permissions.
In particular, as new platforms allow users to revoke granted permissions, to avoid unexpected program behaviors, 
the Android documentation~\cite{Android_document} suggests that apps should perform \textit{Step~2} to check whether they have the required permissions whenever they intend to invoke any permission-protected API.

\subsection{Permission Specification and Its Evolution}
\label{sec:Specification}
On new platforms, dangerous permissions are organized as \textit{permission groups}~\cite{barrera2010methodology}. 
Each dangerous permission in a permission group corresponds to a set of permission-protected APIs~\cite{Android_document}.
As described in Section~\ref{sec:Request Process}, before invoking a permission-protected API, an app needs to request the required permissions following the suggested steps.

However, the Android documentation~\cite{Android_document} does not provide readily available mappings between APIs and their required dangerous permissions.
We call such mappings \textbf{\textit{API-DP mappings}} in our discussion.
The lack of precise API-DP mappings may cause inconvenience to app developers~\cite{discussions}.
Even worse, as the Android platform evolves, API-DP mappings continually change.
The impact of the mapping changes could be huge.
For example, according to the dataset of an existing study~\cite{backes2016demystifying}, there are 22 API-DP mapping changes between API level 23 (Android 6.0) and API level 24 (Android 7.0).
We searched on GitHub and found that the concerned 22 APIs are used by 6,284 apps (see more discussions in Section~\ref{sec:Potential}).
To ensure the compatibility of these apps, the developers may need to exhaustively test the apps on different Android versions and carefully adapt the apps to comply with the evolving permission specification.
Unfortunately, such testing and adaptation are often impractical for app developers.
As a result, many ARP issues, induced by the permission specification evolution, have occurred in real-world Android apps (e.g., issue \#72~\cite{issue72} and issue \#88~\cite{issue88} in {\mycode Secure-preferences}), as we will discuss later.

\section{Characterizing ARP Issues}
\label{sec:empirical}
To understand ARP issues in real-world Android projects, we performed an empirical study.
This section presents the methodology of our study and discusses our findings.



\begin{figure}[t!]
	\centering
	\vspace{-3mm}
	\includegraphics[width=0.28\textwidth]{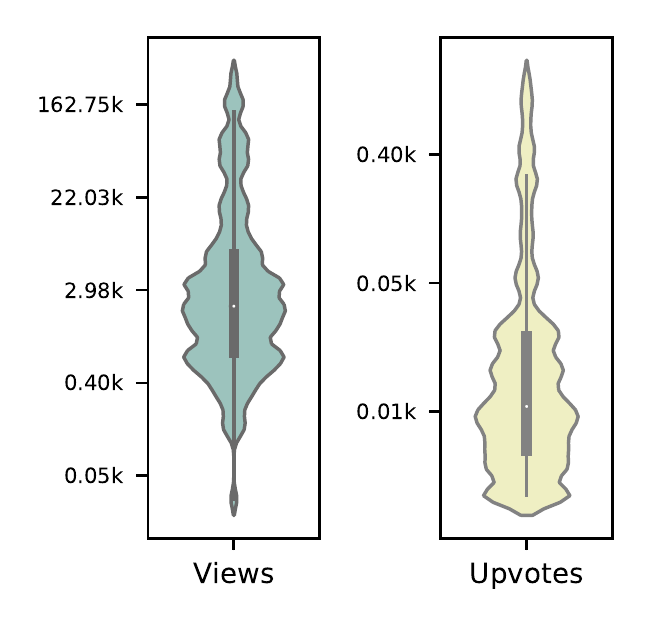}
	\vspace{-2mm}
	\caption{The demographics of the 135 SO posts (in log scale)}
	\label{Posts}
	\vspace{-4mm}
\end{figure}

\subsection{Methodology}
Figure~\ref{fig_overview} provides an overview of our empirical study.
To answer RQ1\textendash RQ3, we need to study the ARP issues in real-world Android projects.
However, when we conducted our study, there were neither readily available datasets of such issues nor precise keywords that can help collect ARP issues from open-source projects.
To bootstrap our research, we resorted to developers' discussions on Stack Overflow~\cite{SO} (SO for short), the most popular question and answer site for professional programmers and enthusiast.
We browsed all SO posts that discuss ARP issues. 
Via analyzing these posts and referring to the official Android documentation, we observed various types of ARP issues and built a taxonomy according to the root causes of the issues.
For each type of issues, we then formulated search keywords and searched the issue trackers of representative Android projects on GitHub for similar issues.
This helped us build a large dataset of real-world ARP issues with buggy code and patches. 
Via studying the issues in this dataset, we further gained a deeper understanding of ARP issues' pervasiveness, seriousness, as well as fixing strategies.
In the remainder of this subsection, we present our study methodology in detail.


\subsubsection{Collecting SO Posts}
As abovementioned, we found no available datasets of ARP issues. 
Therefore, we collected a series of SO posts that discuss ARP issues.
We resorted to SO data instead of open-source projects to bootstrap our research for two reasons.
First, although there are thousands of open-source Android apps with issue trackers on various project hosting sites (such as GitHub), there are no precise search keywords to help retrieve relevant issues.
Simply searching using general keywords like ``permission'' will return a large amount of issues containing substantial noises, which are hard to filter out.
Second, SO is the most popular Q\&A site for developers to share their experience in solving programming issues and SO data have been used for various software engineering research~\cite{linares2014api, duijn2015quality, zhou2020bounties}.
The posts on SO are often tagged manually to categorize questions, which can facilitate our task of searching for developers' discussions on ARP issues.
The reward system of SO also allows us to identify the solutions or explanations of real ARP issues from accepted answers with sufficient confidence.

To collect relevant posts, we searched SO using the \emph{``runtime-permission''} tag and the general phrase \emph{``Android runtime permission''}.
The search returned 2,907 posts.
Next, we removed the duplicates and noises from the search results and kept those posts that satisfy the following three criteria. 
First, the programming issues discussed in the posts should be related to Android runtime permissions.
Second, the posted question should have accepted answers, which are confirmed solutions or explanations.
Third, the posts should have attracted more than 200~\textit{views} or 5~\textit{upvotes} (i.e., having received attention).
With the above process, we obtained 135 high-quality posts.
Figure~\ref{Posts} shows the demographics of these posts.
On average, the posts have attracted 13,919 views (median: 1,657) 
and their associated answers have received 56 upvotes (median: 6). 


\begin{table*}[t!]
	\footnotesize
	\caption{The statistics of the subjects collected in our study}
	\bgroup
	\setlength\tabcolsep{1.0pt}     
	\def\arraystretch{1.2}
	\begin{tabular}{c|c|c|c|c|c|c|c|c|c|c|c|c|c|c|c|c|c|c|c|c|c|c}
		\toprule
		\multirow{2}{*}{} & \multicolumn{3}{c|}{\textbf{\# Stars}} & \multicolumn{3}{c|}{\textbf{\# Forks}} & \multicolumn{3}{c|}{\textbf{\# Downloads}$^\dag$} & \multicolumn{3}{c|}{\textbf{KLOC}} & \multicolumn{3}{c|}{\textbf{\# Commits}} & \multicolumn{3}{c|}{\textbf{\# Issues}} & \multicolumn{3}{c|}{\textbf{\# Permissions}} & \multirow{2}{*}{\textbf{\# Categories}} \\ \cline{2-22}
		& Min.     & Max.     & Avg.    & Min.     & Max.     & Avg.    & Min.      & Max.      & Med.      & Min.    & Max.    & Avg.    & Min.      & Max.      & Avg.      & Min.     & Max.     & Avg.     & Min.       & Max.       & Avg.      &                                \\ \hline \hline
		\textbf{Apps}              & 116        & 14,185        & 1,403       & 31        & 3,587        & 278       & 500+         & 5$\times 10^9$+         & 1$\times 10^5$+         & 4.0      & 961.9      & 62.5      & 33        & 60,631        & 2,201        & 17       & 8,661       & 572       & 1         & 25         & 11        & 30                             \\ \hline
		\textbf{Libraries}         & 53       & 39,467       & 6,061      & 27       & 9,816       & 1,355      & -        & -        & -        & 3.0      & 932.9      & 39.7      & 17         & 22,672         & 970         & 11        & 8,192        & 551        & 1          & 15          & 7        & -                            \\ \bottomrule
		\multicolumn{23}{l}{$^\dag$We give the median downloads of the subjects instead of precise average value, as Google Play only provides the apps' rough download counts.} \\[-0.04cm] 
	\end{tabular}
	\egroup
	\vspace{-3mm}
	\label{tab}
\end{table*}

\begin{figure*}[t!]
	\centering
	\includegraphics[width=0.8\textwidth]{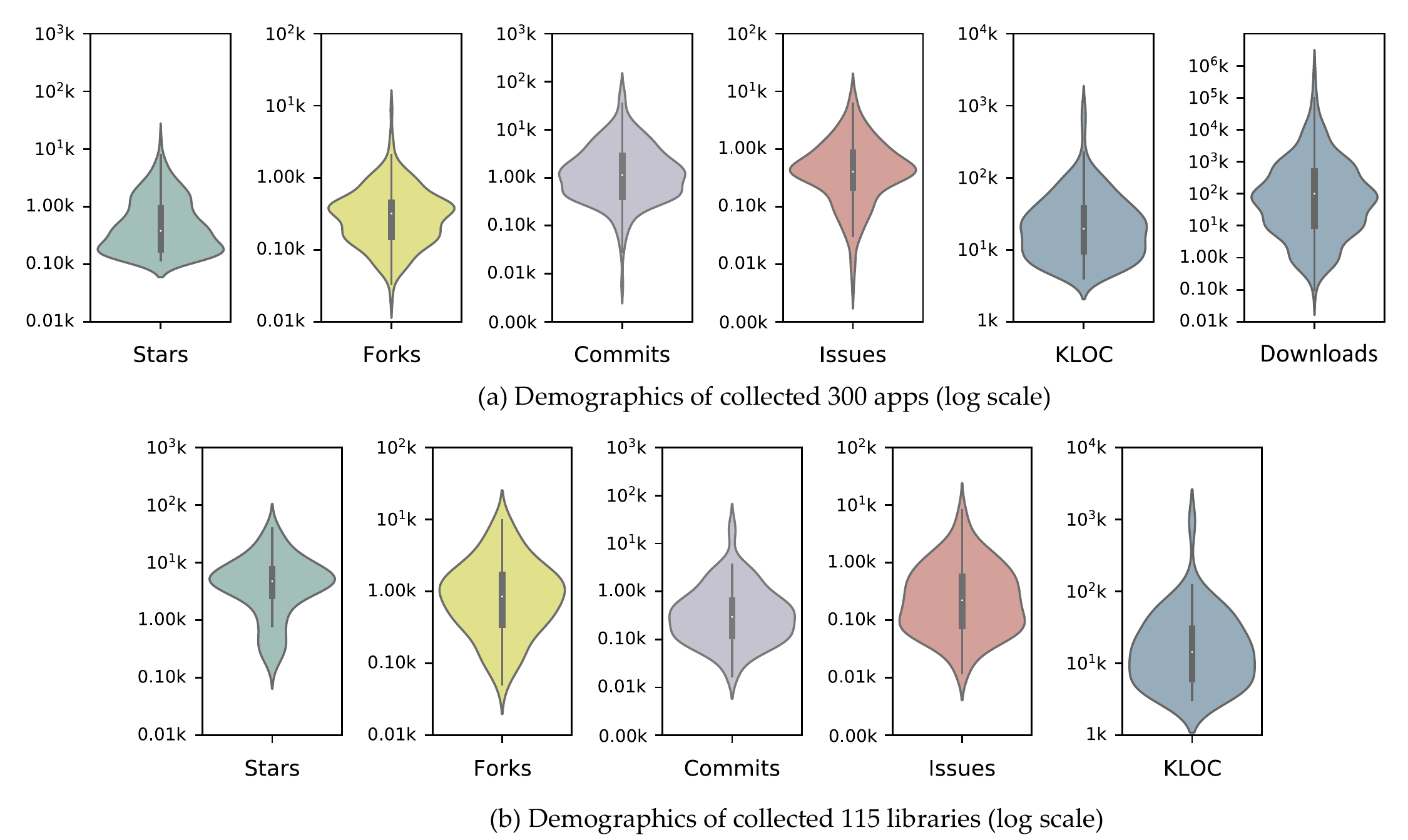}
	
	\caption{Demographics of collected 415 subjects}
	\label{Subjects}
	\vspace{-4mm}
\end{figure*}

\subsubsection{Collecting Open-Source Projects} 

Although SO data can help us address RQ1, the posts are rarely linked to buggy code or patches, which are required to investigate how developers deal with ARP issues at the code level.
Due to this reason, we also collected open-source projects to study ARP issues.
We first crawled the latest version of all open-source apps from F-Droid~\cite{fdroid} (we used the snapshot on 1 Sep 2020) that satisfy the following three criteria.  
First, the apps should declare at least one dangerous permission in their {\mycode manifest} files.
Second, the apps should have publicly accessible issue trackers and code repositories on GitHub. 
We only considered GitHub because it is currently the most popular site to host open-source projects.
Third, the apps should be popular.
To measure popularity, we checked an app's star and fork counts on GitHub, as well as the download counts on Google Play store~\cite{Google_Play}, which is the official app store for Android.
An app is considered popular if it ranks in the top 30\% of all apps indexed by F-Droid according to any of the three metrics.
In addition to apps, we also collected library projects that declare dangerous permissions from GitHub using the keywords \emph{``Android library''} because we observed that many ARP issues are actually introduced by third-party libraries when filtering SO posts.
Similar to apps, we only collected popular libraries that rank in the top 100 according to star or fork counts.
Via the above process, we obtained 300 apps and 115 libraries as the subjects for our study.

Table~\ref{tab} reports the demographics of the 415 subjects and Figure~\ref{Subjects} shows their distributions, including: (1) number of stars, (2) number of forks, (3) number of downloads on {\mycode Google Play} store, (4) thousand lines of code, (5) number of code revisions, (6) number of issues reported in their issue trackers, and (7) number of dangerous permissions registered in their manifest files. 
As we can see from the table, the collected subjects are popular (on average achieving 3,732 stars), well-maintained (on average having 1,586 code commits), diverse (apps belong to 30 different categories), and prone to ARP issues (on average declaring 9 dangerous permissions).
In particular, among the collected 300 apps, 216 of them are also released on Google Play store, and 32 out of the above 216 apps are ranked in the top 500 in 13 categories.
Such demographics demonstrate the representativeness of our collected subjects.
Besides, all of these open-source projects were started before 2015, the year that Google introduced the runtime permission mechanism in Android OS.
This enables us to analyze the impacts of the evolution of runtime permission mechanism.

\vspace{-2mm}
\subsubsection{Data Analysis}
This subsection briefly describes how we performed data analysis to answer RQ1\textendash RQ3.
More details will be given in subsequent sections.

To answer \textbf{RQ1}, we performed an in-depth analysis of the 135 SO posts.
Specifically, we followed an open coding procedure~\cite{open-coding}, a widely-used approach for qualitative research, to categorize the issues discussed in each post.
Initially, based on the knowledge acquired from the data collection and post filtering process, we determined a draft issue taxonomy and discussed how to label the issues mentioned in the posts according to their symptoms and root causes.
Then, three authors (annotators) of this paper, who all have more than three years of Android development experience, independently analyzed each post, including the issue descriptions, comments, accepted answers, and linked resources (e.g., relevant issues on GitHub).
They also referred to the Android documentation to aid the understanding of ARP issues.
After the first round of independent analysis and labeling, the three annotators gathered to compare and discuss their results, in order to clarify the descriptions and boundaries of different categories, add and remove labels, and adjust the hierarchical structures of categories.
This led to a more clear-cut labeling strategy.
Next, the three annotators continued the labeling process and iteratively refined the results as well as the labeling strategy.
When there were conflicting opinions during the iterations, the other authors were invited to meetings to resolve conflicts.
The annotators finally reached a consensus on the issue taxonomy after six iterations. 

To answer \textbf{RQ2}, we first formulated keywords for each type of ARP issues observed from analyzing the SO posts to search for similar issues in the 415 GitHub projects.
This step resulted in a collection of 199 real ARP issues with detailed information, including the issue reports, the buggy code, and the patches.
We then studied the issue distribution to understand the pervasiveness of each issue type and studied the evolutionary trends of the issues by analyzing the yearly data.
We also paid special attention to the issue types that have become increasingly common in recent years and investigated their potential impact by looking at how many open-source projects may suffer from such issues.

To answer \textbf{RQ3}, we studied the patches of the issues collected from the 415 GitHub projects to understand how developers fixed the issues and distill common strategies.

\subsection{RQ1: Issue Types and Causes}
\label{sec:RQ1}


\begin{figure*}[t]
	\centering
	\includegraphics[width=0.9999\textwidth]{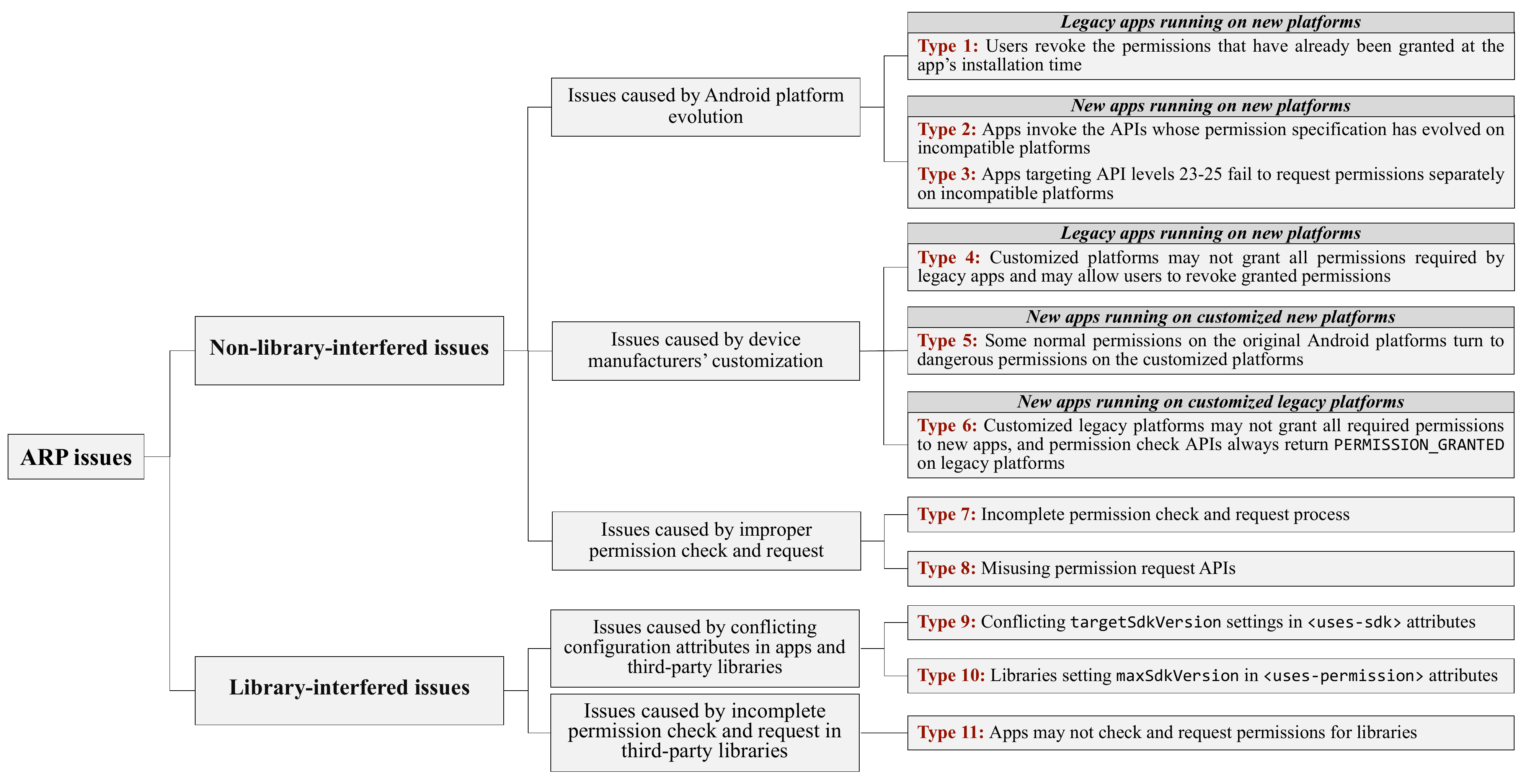}
	\caption{The taxonomy of our studied ARP issues}	
	\label{fig_tree}
\end{figure*}

\tcbset{colback=gray!5!white, colframe=black,
	notitle, 
	width={\linewidth+1pt},
	top=1pt,
	left=1pt,
	right=1pt,
	bottom=1pt,
	toprule=0.5pt,
	titlerule=1pt,
	bottomrule=0.5pt,
	leftrule=0.5pt,
	rightrule=0.5pt,
	after skip=6pt,}
\vspace{4pt}
\noindent\begin{tcolorbox}
	\small
	\noindent\textbf{Finding 1:} \emph{ARP issues not only can be caused by Android apps' mishandling of dangerous permissions, but also can occur due to the interference of third-party libraries.}
\end{tcolorbox}

We observed that our studied ARP issues can be classified into two general categories: \textit{\textbf{non-library-interfered issues}} and \textit{\textbf{library-interfered issues}}.
The former are caused by the mishandling of dangerous permissions in app modules while the latter are introduced by inconsistent permission handling between app modules and third-party libraries.
These two categories of issues can be further classified into 11 subcategories according to the root causes of the issues (the 11 issue types are disjoint).
Figure~\ref{fig_tree} summarizes all issue types observed by us.
To the best of our knowledge, the figure presents by far the most comprehensive categorization of ARP issues.\footnote{The APR issues discussed in this section are not specific to dangerous permissions. Some may also occur when dealing with normal permissions.}
In the following, we discuss each type of issues in detail. 

\subsubsection{Non-Library-Interfered ARP Issues}
\label{sec:Pattern A}

\vspace{4pt}
\noindent\begin{tcolorbox}
	\small
	\noindent\textbf{Finding 2:} \emph{ARP issues can be caused by the evolution of the Android platform, in particular, the permission mechanism. It is hard for developers to track the changes of the permission mechanism across various Android versions and continuously adapt their apps to avoid potential ARP issues.}
\end{tcolorbox}

As discussed earlier, on Android platforms prior to Marshmallow (6.0), apps could be granted all requested dangerous permissions at installation. The permissions, once granted, cannot be revoked by users.
In such scenarios, apps rarely suffer from permission-related issues if the developers carefully declare all required permissions in the {\mycode manifest} files.
However, with the adoption of the runtime permission model, ARP issues arise in more and more apps.
We observed three typical types of such issues, which occur in two different scenarios.

\textbf{Scenario 1: Legacy apps running on new platforms.} For backward compatibility, new platforms allow legacy apps to be used by granting all dangerous permissions declared in the apps' {\mycode manifest} file.
However, on new platforms, users can freely revoke the permissions granted to apps.
If permission revoking happens to legacy apps, the {\mycode SecurityException}s will be thrown whenever the apps invoke permission-protected APIs (\textbf{Type~1 issues}).
A typical example is issue \#1343~\cite{issue1343} in the app {\mycode Owncloud}.
As discussed in SO post \#35248269~\cite{PatternAa1}, developers of such legacy apps have no choice but to migrate their apps to use the runtime permission mechanism.
Otherwise, the apps will very likely receive poor user ratings, since they can easily crash on new platforms.

\textbf{Scenario 2: New apps running on new platforms.}
Unlike legacy apps, new apps request dangerous permissions at runtime.
Ideally, if the developers carefully implement the permission check and request process, these apps should work fine on new platforms.
However, Android platform evolves quickly~\cite{Chang2020}. 
We observed that new apps may suffer from ARP issues due to such evolution.
Particularly, we observed the following two types of issues:

\begin{itemize}
	\item \emph{\textbf{Evolution of API-DP mappings.}}
	ARP issues may arise due to the evolution of API-DP mappings across different Android versions, including: 1) the addition/deletion of permission-protected APIs, 2) the addition/deletion of dangerous permissions, and 3) the changes of associations between APIs and permissions. 
	As discussed in SO post \#20740632~\cite{discussions}, app developers can hardly realize the above changes. 
	What is more, there are no official documentations providing the precise API-DP mappings for each Android version and highlighting the changes across versions~\cite{dawoud2021bringing, aafer2018precise}. 
	Therefore, if an app uses an API whose permission specification has undergone changes, it may crash when invoking such an API on certain incompatible platforms (\textbf{Type 2 issues}). 
	An example is issue \#982~\cite{issue982} in {\mycode Lawnchair}.
	
	\item \emph{\textbf{Evolution of the permission group mechanism.}}
	The new platforms with API levels 23\textendash25 handle runtime permission requests at the group level.
	On such platforms, if a user chooses to grant a permission requested by an app, the rest dangerous permissions in the same group that are declared in the app's {\mycode manifest} file will also be granted, without the need of further user interactions.
	However, this mechanism has changed since Android 8.0 (API level 26).
	For apps targeting Android 8.0 or later versions, they are granted only the permissions that are explicitly requested at runtime.
	The other permissions in the same group as the requested permissions should still be requested separately.
	As discussed in SO post \#47787577~\cite{post47787577}, such evolution complicates the use of dangerous permissions.
	Moreover, it can cause app crashes if developers do not carefully program their apps to request each permission separately on Android 8.0 and above (\textbf{Type 3 issues}).
	An example is issue \#58~\cite{issue58} in {\mycode Buglife}. 
	
\end{itemize}

\vspace{6pt}
\noindent\begin{tcolorbox}
	\small
	\noindent\textbf{Finding 3:} \emph{
		Permission-related system customizations made by Android device manufacturers may also induce ARP issues.
		Such issues occur only on specific device models and are difficult to debug.
	}
\end{tcolorbox}
\vspace{2pt}

Android is an open-source mobile OS. Many mobile device manufacturers build their products by customizing the original Android platform.
The customization may also affect the permission mechanism and cause ARP issues.
We observed that such customization-induced ARP issues may occur in three different scenarios:

\textbf{Scenario 1: Legacy apps running on customized platforms.}
As discussed in SO post \#39654809~\cite{PatternAb1}, customized Android platforms, either new or legacy, may not grant all the requested permissions to legacy apps and may allow users to revoke granted permissions.
If legacy apps cannot obtain the required permissions when running on such customized platforms, the {\mycode SecurityException}s will be thrown when the apps invoke permission-protected APIs (\textbf{Type 4 issues}).
Take issue \#261~\cite{issue261} in {\mycode NEMAndroidApp} for example.
Developers complained that when the app runs on Oppo R7 (with Android 4.4.4), the system's built-in security software would block the requests for dangerous permissions, which led to app crashes.

\textbf{Scenario 2: New apps running on customized new platforms.}
Due to customization, some normal permissions on the original Android platform may turn into dangerous permissions on the customized platforms.
This could lead to app crashes since the majority of apps would not request for such permissions (\textbf{Type 5 issues}).
For instance, according to SO post \#59418504~\cite{PatternAb2}, some Xiaomi devices with API level 26 treat the ``{\mycode Display pop-up window}'' permission, which is a normal permission on the original Android platform, as a dangerous permission. 
This problem caused many app crashes and similar cases frequently happened, which was troubling to app developers.

\textbf{Scenario 3: New apps running on customized legacy platforms.}
Customized legacy platforms may not grant all requested permissions to new apps and may allow users to revoke granted permissions.
When new apps run on such platforms, they will check and request for dangerous permissions at runtime.
For better compatibility, they may use the API {\mycode ContextCompat.checkSelfPermission()} provided by the Android support library for permission check. 
However, this API always returns {\mycode PERMISSION\_GRANTED} on legacy platforms.
The new apps then may proceed to invoke permission-protected APIs.
As discussed in SO post \#43467522~\cite{PatternAb3}, if the required permissions of these APIs are actually not granted or have been revoked, the apps will crash due to {\mycode SecurityException}s (\textbf{Type 6 issues}).
An example is issue \#125~\cite{issue125} in {\mycode QRCodeReaderView}.

\vspace{4pt}
\noindent\begin{tcolorbox}
	\small
	\noindent\textbf{Finding 4:} \emph{ 
		Developers may make mistakes in the permission check and request process, causing various ARP issues.
	}
\end{tcolorbox}

As mentioned in Section~\ref{sec:Request Process}, a proper permission check and request process involves four steps.
We noticed that this is a non-trivial task for many developers and the mistakes they make could cause two different types of ARP issues:

\textbf{Incomplete permission check and request process.} ARP issues would arise when developers miss certain steps in checking and requesting for dangerous permissions (\textbf{Type 7 issues}).
Specifically, we observed the following cases:

\begin{itemize}[topsep=1pt]
	
	\item Developers may forget to declare the required dangerous permissions in their app's {\mycode manifest} file (i.e., missing step 1).
	As discussed in SO post 44374646~\cite{PatternAa2i1}, in such cases, the permission request dialog would not pop up after calling the Android API {\mycode ActivityCompat.requestPermissions()} and the permissions cannot be successfully granted.
	As such,  the app would crash when the APIs that require the permissions are invoked at runtime.
	An example is issue \#229~\cite{issue229} in {\mycode Dexter}.
	
	\item Developers may not be fully aware of the permission specification of Android APIs and may only declare the required dangerous permissions in their app's {\mycode manifest} file without really making the app to request the permissions at runtime (i.e.,missing steps 2\textendash4).
	As discussed in SO post 34959331~\cite{PatternAa2i2}, this 
	will also cause app crashes~\cite{PatternAa2i2}.
	An example is issue \#1~\cite{issue1} in {\mycode Swati4star}. 
	
	\item Developers may also forget to handle users' response to permission requests (i.e., missing step 4).
	In this scenario, the subsequent operations that require permissions may not be properly performed, according to the SO post 42762308~\cite{PatternAa2i3}. 
	An example is issue \#737~\cite{Issue737} in {\mycode Native-navigation}. 
	
	\item Android documentation suggests that apps should explain to users why they need certain dangerous permissions~\cite{notes}. 
	As discussed in SO post 30719047~\cite{PatternAa2i4}, if there is no proper explanation, users may deny the permission requests or even check the option \textit{``Never ask again''} on the permission request dialog.
	Then the operations that require the permissions may never be performed properly.
	An example is issue \#10~\cite{issue10} in {\mycode Assent}.
\end{itemize}

\textbf{Misusing permission request APIs.} 
Android provides a series of APIs to request permissions.
We observed that developers sometimes use wrong APIs for permission requests, causing ARP issues (\textbf{Type 8 issues}).
For example, as discussed in SO post 35989288~\cite{Apost}, the {\mycode Activity} class (for creating a window) and {\mycode Fragment} class (for creating a portion of a window) define their own permission request methods.
If an app misuses these methods, such as calling the permission request method defined in {\mycode Activity} class from a {\mycode Fragment}, the permissions cannot be granted properly.
An example is issue \#228~\cite{issue228} in {\mycode OpenRedmine}.	

\subsubsection{Library-Interfered ARP Issues}

\vspace{4pt}
\noindent\begin{tcolorbox}
	\small
	\noindent\textbf{Finding 5:} \emph{
		ARP issues may arise due to the conflicts between the configuration attributes in the {\mycode manifest} file of apps and third-party libraries.
	}
\end{tcolorbox}
\vspace{2pt}

For code reuse, Android apps often reference third-party libraries.
Many libraries contain {\mycode manifest} files to present essential information to the Android platform.
However, as apps and libraries are typically maintained by different developers, the configuration attributes declared in an app's {\mycode manifest} file may have conflicts with those declared in a library's {\mycode manifest} file.
In such cases, Android build tool would resolve the conflicts following predefined rules and merge multiple {\mycode manifest} files into one~\cite{Android_document}.
Such merging often causes the following two types of ARP issues.

\begin{itemize}
	\item \textbf{Conflicting {\mycode targetSdkVersion} settings in {\mycode <uses-sdk>} elements.}
	The {\mycode <uses-sdk>} element in the {\mycode manifest} file of an app/library specifies the API levels on which the app/library is able to run.
	Among them, the {\mycode targetSdkVesion} attribute is often set to indicate the best-fit API level of the app/library.
	If the {\mycode targetSdkVesion} attribute declared in an app is in conflict with that declared in any of the app's libraries, Android built tool  will choose the app's configuration over the library's to resolve conflicts.
	As discussed in SO post 58059097~\cite{PatternbBa1}, in such cases, if the library, which is forced to run on an unintended API level, does not properly check and request for permissions before invoking certain APIs (note that these APIs may not require permission protection on the library's target Android version), there will be runtime exceptions that can cause app crashes (\textbf{Type 9 issues}).
	An example is issue \#920~\cite{issue920} in {\mycode Android-beacon-library}.
	
	
	\item \textbf{Libraries' setting of {\mycode maxSdkVersion} in {\mycode <uses-permission>} elements.}
	The {\mycode <uses-permission>} elements in the {\mycode manifest} file of an app/library are used to specify the permissions that the user must grant to ensure the app/library to operate correctly.
	For this element, the {\mycode maxSdkVesion} attribute is often set to indicate the highest API level on which the permission should be granted to the app\footnote{Setting {\mycode maxSdkVersion} is useful if the permission an app requires is no longer needed beginning at a certain API level~\cite{Android_document}.}.
	When an app and its referenced library declare a same dangerous permission in their {\mycode manifest} files but only the library specifies a {\mycode maxSdkVersion}, the final {\mycode manifest} file merged by the Android build tool will contain the dangerous permission with the {\mycode maxSdkVersion} set by the library.
	Then the app cannot request for the concerned permission on the platforms whose API levels are greater than {\mycode maxSdkVersion}.
	As discussed in SO post 54615936~\cite{PatternbBa2}, the app will crash in such cases (\textbf{Type 10 issues}).
	An example is issue \#723~\cite{issue723} in {\mycode Mapsforge}.
	
\end{itemize}

\vspace{6pt}
\noindent\begin{tcolorbox}
	\small
	\noindent\textbf{Finding 6:} \emph{Third-party libraries may invoke permission-protected APIs without checking whether the required permissions are granted. If apps using these libraries do not properly check and request for the required permissions, ARP issues will arise.
	}
\end{tcolorbox}
\vspace{2pt}

To access user data or system features, third-party libraries often need to invoke permission-protected APIs.
Since libraries typically are not supposed to deal with user interactions directly, many library developers would assume that it is the apps' job to check and request for the permissions required by libraries.
However, as discussed in SO post 36693127~\cite{discussions12}, app developers may not do so and this would cause {\mycode SecurityException}s (\textbf{Type 11 issues}).
As we will show later, such issues are quite common, since app developers may not be aware of the need to check and request for the permissions required by libraries.
To avoid such ARP issues, in SO  post~39603098~\cite{discussions11}, a developer suggested that an app should manage all the dangerous permissions it directly or transitively requires, rather than expecting the libraries to do so. 

\subsection{RQ2: Pervasiveness and Seriousness}
\label{sec:RQ2}

Section~\ref{sec:RQ1} presents 11 types of ARP issues we observed via studying SO posts and relevant issues on GitHub.
In this section, we further study how pervasive and serious these issues are in practice.
To answer RQ2, we first refine the keywords to search for more instances of the 11 types of ARP issues on GitHub, and then analyze the yearly distribution and evolution trends of the collected issues.
In the following, we present our study approach in detail and then discuss our findings.

\textbf{Keyword formulation.} For each type of ARP issues in our taxonomy, we formulated search keywords, as listed in Table~\ref{t2}, to look for more instances in the 415 collected GitHub projects.
In the table, the symbol ``$\times$'' denotes Cartesian product and ``$\wedge$'' means ``and''.
The keywords for issues of Types 1, 3, and 7\textendash11 are those terms that frequently occur in our studied SO posts and relevant issue reports. 
Issues of Types 4\textendash6 are caused by device manufacturers' customization of the original Android platform.
To search for such issues, we included the names of six top Android device manufacturers, which in total have over 80\% market share according to App Brain~\cite{Topdevicebrands}, in our keywords.
Finally, to look for issues of Type 2, we should include the APIs or permissions in the evolved API-DP mappings in the keywords.
However, as discussed in Section~\ref{sec:Specification}, there are no readily available mappings between Android APIs and dangerous permissions in the Android documentation~\cite{Android_document}. 
Although existing studies (e.g., ~\cite{felt2011android, au2012pscout, bartel2012automatically, bartel2014static, nguyen2017android, backes2016demystifying}) proposed several techniques to infer such mappings, we found that their released datasets are outdated.
To precisely locate issues of Type~2, we chose to infer the mappings by ourselves.
As pointed out by Bogdanas~\cite{bogdanas2017dperm}, starting from Android 6.0 (API level 23), Google formally documents permission specifications in two ways: 1) using Java annotation {\mycode @requiresPermission} to associate APIs with permissions and 2) using {\mycode @link android.Manifest.permission\#} to describe an API's required permissions.
Inspired by this observation, we developed a tool named \textsc{APMiner} to analyze the Javadoc of the Android SDK~\cite{AndroidSDK} to infer API-DP mappings.\footnote{APMiner is available at http://arp-issues.github.io/. The tool is not a major contribution of this work. We built it for the purpose of mining the latest API-DP mappings.}
We applied \textsc{APMiner} on API levels 23\textendash29 and compared the mined mappings across these API levels.
The names of the APIs and permissions in the evolved API-DP mappings were then included in our keywords.

\textbf{Issue collecting and filtering.} Next, we used the keywords to search the issue trackers of the 415 GitHub projects.
We removed the duplicates and noises from the search results and kept those issue reports that satisfy the following criteria: (1) the reported issue is an instance of our observed types and (2) the report contains sufficient information for us to understand the reported issue. 
With this process, we obtained a collection of 199 real ARP issues.
Note that there might be ARP issues that are not included in our taxonomy (i.e., not instances of the 11 types) but we did not observe such cases in our study.

\subsubsection{The Distribution of ARP Issues}
\label{sec:Distribution}

\begin{table*}[t!]
	\centering
	\footnotesize
	\caption{The search keywords for each issue type}
	\bgroup
	\setlength\tabcolsep{5pt}     
	\def\arraystretch{1.15}
	\vspace{-3mm}
	\begin{tabular}{l|l}
		\toprule
		\textbf{Types} & \textbf{Keywords (case insensitive)}\\ \hline \hline
		1       & \{``API level''\} $\times$ \{19, 20, 21, 22, 23\} $\times$ \{``runtime permission''\}                              \\ \hline
		2 & \textit{API and permission in evolved API-DP mappings} $\times$ \{``runtime permission''\}              \\ \hline
		3, 7, 8     & \{manifest, check, request, handle\} $\times$ \{``runtime permission''\} \\ \hline
		4, 5, 6        & \{Samsung, Huawei, Xiaomi, Oppo, Motorola, Vivo\} $\times$ \{``runtime permission''\} \\ \hline
		9, 10        & ``merge manifest'' $\wedge$ ``runtime permission''                                      \\ \hline
		11         & ``library'' $\wedge$ ``runtime permission''                                             \\ \hline
		\bottomrule
	\end{tabular}
	\egroup
	\label{t2}
\end{table*}

\noindent\begin{tcolorbox}
	\small
	\noindent\textbf{Finding 7:} \emph{71\% of our collected issues are non-library-interfered issues while the remaining 29\% are library-interfered issues. Among the library-interfered issues, those that are induced by the evolution of API-DP mappings (Type~2) are the most common. Among the library-interfered issues, those that arise after merging conflicting {\mycode manifest} files are the most common (Type 11).
	}
\end{tcolorbox}
\vspace{2pt}

\begin{figure}[t!]
	\vspace{-2mm}
	\centering
	\includegraphics[width=0.30\textwidth]{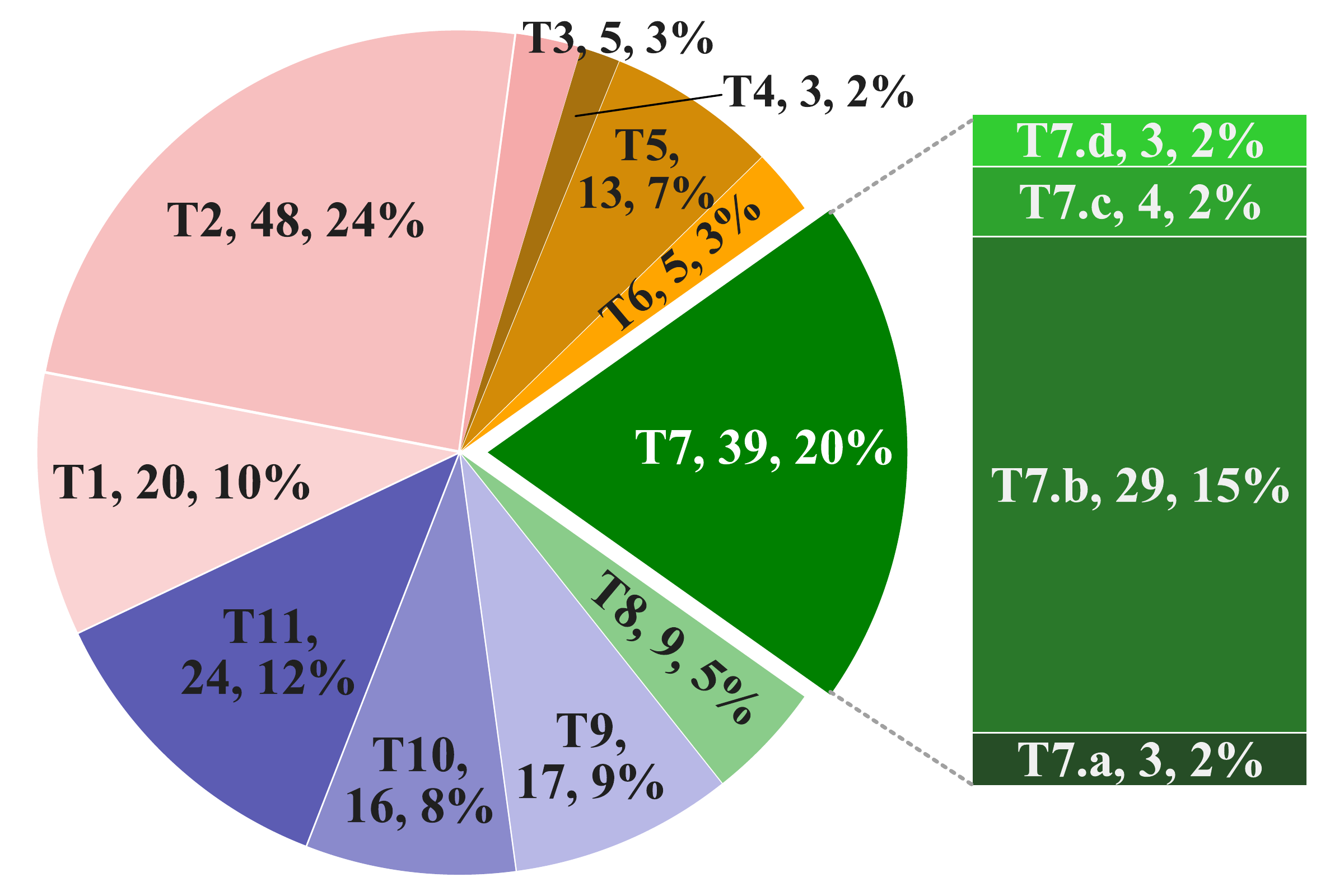}
	\includegraphics[width=0.41\textwidth]{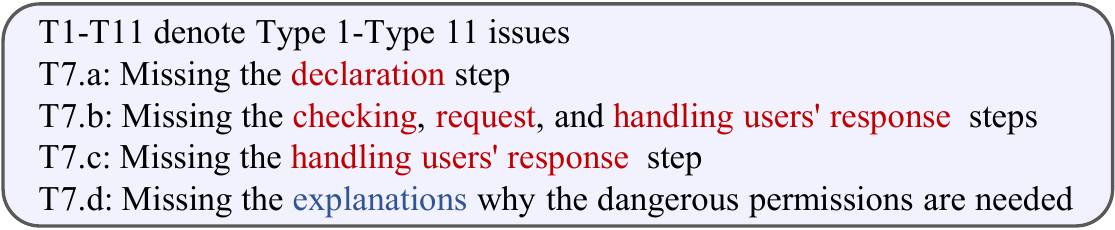}
	\caption{The distribution of ARP issues}	
	\label{fig_pie}
	\vspace{-6mm}
\end{figure}

Figure~\ref{fig_pie} gives the distribution of the 199 issues. We can make the following observations from the distribution:

\begin{itemize}
	\item Among the non-library-interfered issues, 48 (24\%) were induced by the evolution of API-DP mappings (Type 2).
	We conjecture that such issues are the most pervasive because there is no effective mechanism or tool to help developers adapt their apps to the continually-changing API-DP mappings.
	
	\item The second common type of non-library-interfered issues are those caused by the incomplete permission check and request process (Type 7).
	As Figure~\ref{fig_pie} shows, the majority of such issues occurred because the developers did not check and request permissions for certain permission-protected APIs.
	The issues are pervasive, probably because there are many permission-protected APIs on the Android platform. 
	So, it is not easy for developers to check and request permissions properly when their apps use such APIs without automated tool support.
	
	\item Library-interfered issues are also common, accounting for 29\% of all our collected issues.
	Among these issues, 33 (17\%) occurred due to the conflicts between the configuration attributes in the {\mycode manifest} files of apps and third-party libraries (Types 9\textendash10).
	Such issues frequently occurred because apps and libraries are often maintained separately by different parties.
	It is hard for developers to be aware of the configurations of all libraries used by their apps.
	The remaining 24 issues (12\%) occurred because the libraries use certain permission-protected APIs without properly requesting permissions (Type 11).
	Interestingly, app developers and library developers seem to have divergent opinions on which layer is responsible for requesting permissions:
	For four out of the 24 issues, the library developers acknowledged that the permissions should be requested in the libraries;
	However, for the other 20 of the 24 issues, library developers expected app developers to request permissions instead.
\end{itemize}


\subsubsection{The Evolution Trends of ARP Issues}
\label{sec:Evolutionary}

\vspace{6pt}
\noindent\begin{tcolorbox}
	\small
	\noindent\textbf{Finding 8:} \emph{ARP issues caused by third-party libraries (Types 9\textendash11), the evolution of API-DP mappings (Type~2), and device manufacturers' customization of normal permissions (Type~5) are becoming increasingly pervasive in the last five years. In contrast, issues affecting legacy apps that have not migrated to the runtime permission model (Type 1) and those arising from the incomplete permission check and request process (Type 7) are diminishing. 
	}
\end{tcolorbox}
\vspace{2pt}

Our collected 415 GitHub projects all started before Android adopted the runtime permission model.
Observing this, we further analyzed the yearly distribution of the 199 ARP issues to explore the evolution trends of these issues.
We ignored issues of Types 3, 4, 6, and 8 because they only account for a small proportion (less than 5\%) in our dataset.
Analyzing the evolution trend of these issues might result in observations that have little statistical significance. 

\begin{figure}[t!]
	\centering
	\includegraphics[width=0.48\textwidth]{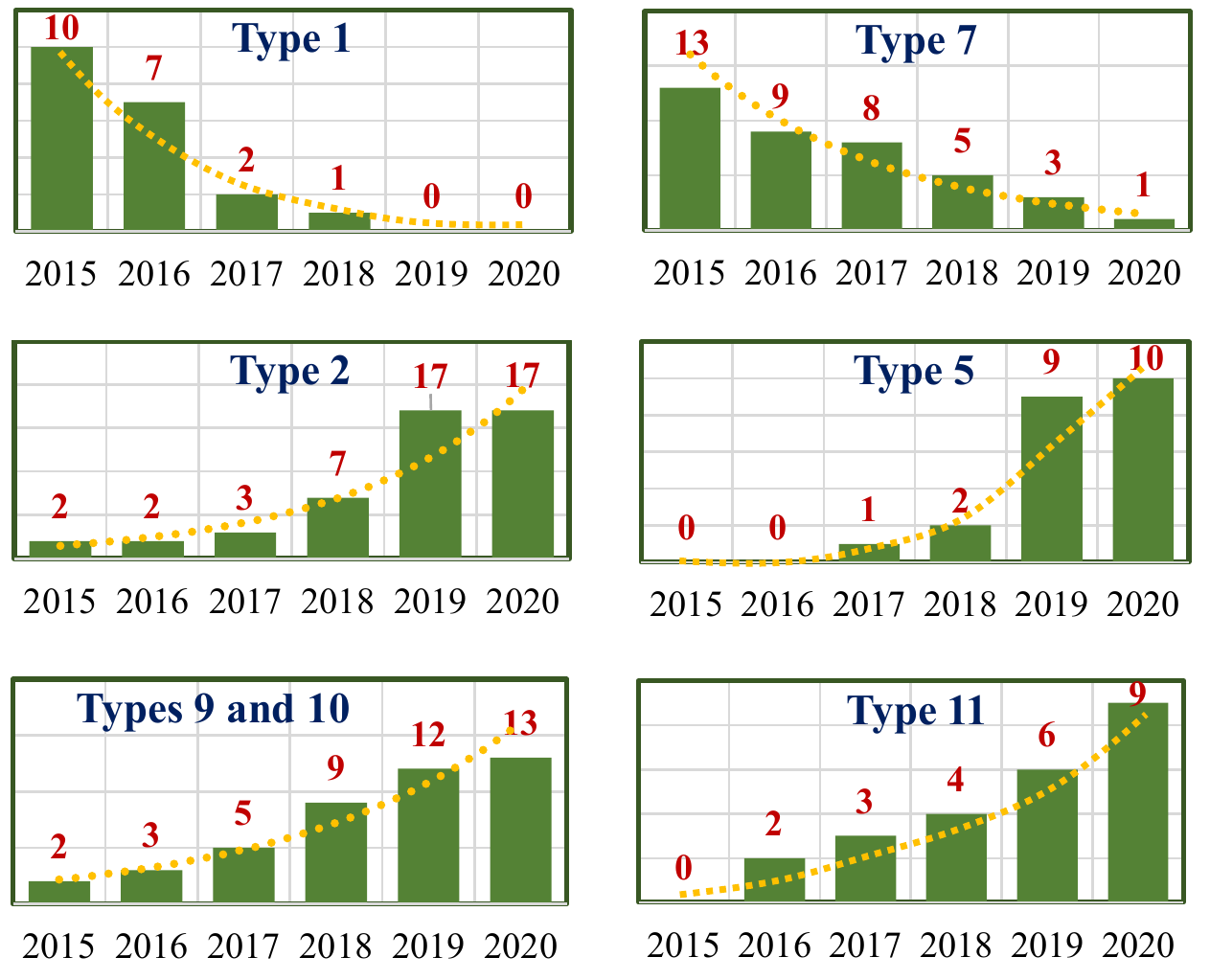}
	\caption{Distribution of ARP issues from 2015 to 2020}	
	\label{fig_time}
\end{figure}

Figure~\ref{fig_time} gives the yearly distribution for issues of Types 1, 2, 5, 7, and 9\textendash11. From the figure, we can make the following observations:

\begin{itemize}
	\item Issues of Types 1 and 7 gradually diminish over the years.
	These issues occurred because developers did not migrate their apps to the runtime permission model or did not check and request permissions properly. Since the runtime permission model has been used for years, it is natural that these issues are fading away and becoming less relevant.
	\item On the other hand, issues of Types 2 and 5 are increasingly pervasive. These issues occurred due to the evolution of API-DP mappings and Android device manufacturers' customization of normal permissions, respectively. This finding suggests that the evolution and customization of Android platforms may affect the compatibility of apps, resulting in serious runtime permission issues.
	\item The number of library-interfered issues (Types 9\textendash11) continually increases over the years. 
	To understand why, we analyzed the build scripts of our collected open-source apps.
	We found that the number of libraries used by these apps on average increased by nearly 35\% across the latest ten releases.
	Such intensive uses of third-party libraries have likely induced various ARP issues, as libraries may interfere with apps' handling of dangerous permissions.
\end{itemize}

\subsubsection{Potential Seriousness of ARP Issues}
\label{sec:Potential}
\vspace{6pt}
\noindent\begin{tcolorbox}
	\small
	\noindent\textbf{Finding 9:} \emph{
		ARP issues of Types 2, 9, and 10 might be very serious in real-world Android projects. Without automated tool support, developers can hardly realize such issues.
	}
\end{tcolorbox}
\vspace{0.2em}

\begin{figure*}[t!]
	\centering
	\includegraphics[width=0.88\textwidth]{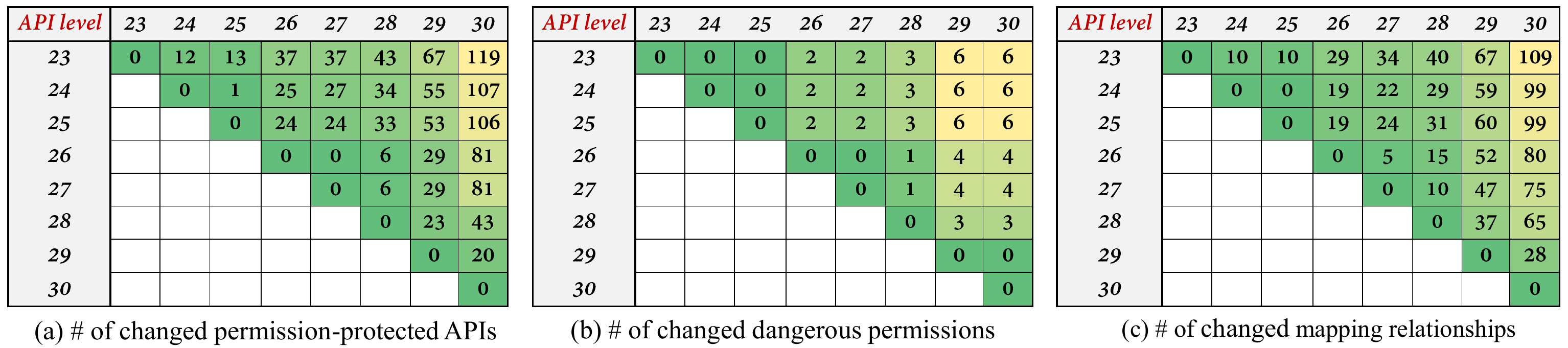}
	\caption{The evolution of API-DP mappings across different Android versions (pairwise comparisons)}	
	\label{fig_mappings}
\end{figure*}
As discussed in Section~\ref{sec:Evolutionary}, ARP issues of Types 2, 5, and 9\textendash11 are becoming increasingly pervasive in recent years.
To estimate how seriously these issues may affect the Android ecosystem, we performed further analysis, leveraging larger-scale data.

Type 2 issues are caused by the evolution of API-DP mappings.
To understand the potential seriousness of such issues, we analyzed the changes of our extracted API-DP mappings (see Section~\ref{sec:Distribution}) between different Android versions.
As shown in Figure~\ref{fig_mappings}, there are three types of changes, including: 1) the addition/deletion of permission-protected APIs, 2) the addition/deletion of dangerous permissions, 3) and the changes in the mapping relations between APIs and permissions.
In total, there are over one hundred such changes between API levels 23 and 30, which is not a small number.
More surprisingly, we searched on GitHub and found 16,492 Android projects that use at least one of the evolved APIs or declare at least one of the evolved permissions.
As an example, we observed that the permission-protected API {\mycode TelephonyManager.getDeviceId()}, which is evolved at API level 29, is used by 925 Android projects.
\textit{Since such APIs or permissions are frequently used by a large number of Android projects, ARP issues may easily arise if developers do not carefully deal with the evolution of the APIs/permissions.}

Issues of Types 9\textendash10 are caused by conflicting configurations between apps and third-party libraries.
To understand how serious these issues can be in practice, we randomly sampled 3,487 apps, which use Gradle as the build tool, on GitHub, and extracted their referenced libraries by analyzing the {\mycode build.gradle} files.
Alarmingly, we found that 2,946 apps (84.5\%) depend on at least one library that uses inconsistent {\mycode targetSdkVersions}.
We also randomly sampled 1,062 Android libraries on GitHub.
We were again surprised to find that 316 sampled libraries (29.8\%) set {\mycode maxSdkVersions} for their declared dangerous permissions. 
According to \textit{libraries.io}~\cite{Librariesio}, a website for querying the dependencies among projects, these 316 libraries are used by approximately 4,130 Android projects. 
\textit{These projects can easily suffer from issues of Types 9/10, if developers do not carefully resolve the conflicts between the configurations of apps and its referenced libraries.} 

We did not further analyze issues of Type~5 because they concern customized Android platforms, which are mostly closed-source.
We did not further analyze Type~11 issues either, as it is known that many Android apps rely on various libraries, and detecting such issues via analyzing the complex dependencies is not a trivial task.
We will leave this to be explored in our future work.

\subsection{RQ3: Fixing Strategies}
\label{sec:RQ3}

\begin{figure*}[t!]
	\centering
	\includegraphics[width=0.9999\textwidth]{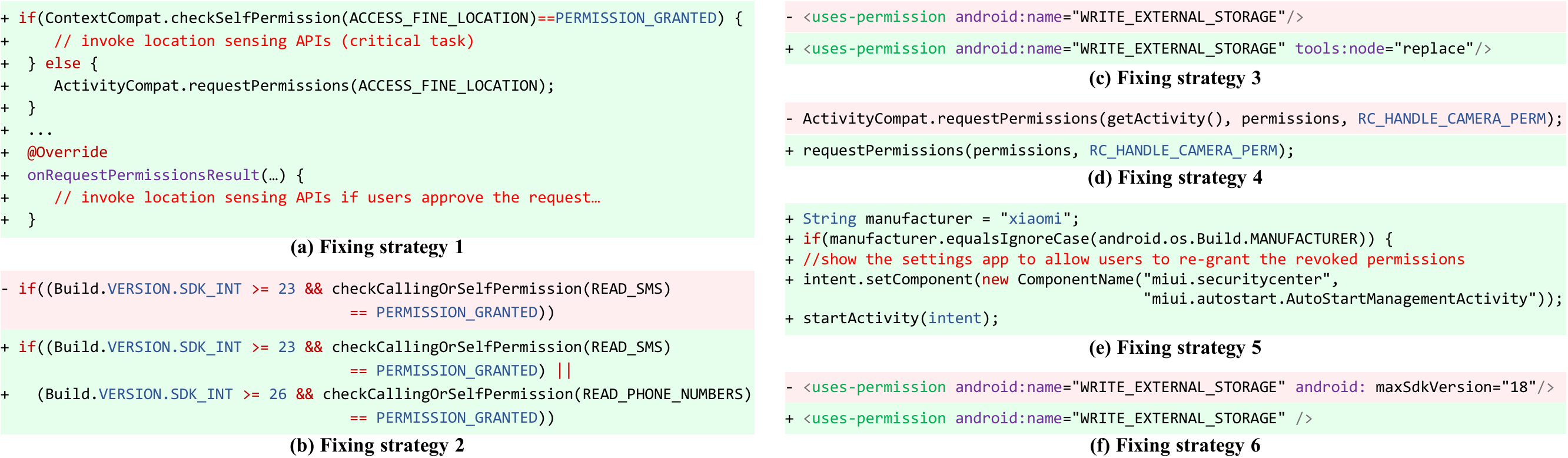}
	\caption{Example patches of fixing strategies (simplified to ease understanding; the code may not be syntactically correct)}
	\label{fig_patch1}
\end{figure*}

To answer RQ3, we first analyzed our collected 199 ARP issues and found that: 1) 135 issues have already been fixed and we can find the corresponding patches, 2) 42 issues are not yet fixed but the developers have described their fixing solutions in the issue reports, and 3) for the remaining 22 issues, we could not find relevant discussion in the issue reports to understand how the issues can be fixed.
Therefore, we further analyzed the 177 issues, for which we could figure out the fixing solutions, to learn the common strategies used by developers to fix ARP issues.

\vspace{6pt}
\noindent\begin{tcolorbox}
	\small
	\noindent\textbf{Finding 10:} \emph{
		We observed six fixing strategies, four of which can be used to resolve multiple types of ARP issues.
	}
\end{tcolorbox}
\vspace{2pt}

\begin{table}[t!]
	\scriptsize
	\caption{Statistics of fixing strategies}
	\bgroup
	\setlength\tabcolsep{2pt}     
	\def\arraystretch{1.2}
	\begin{tabular}{c|c|c|c|c|c|c|c|c|c|c|c}
		\toprule
		\diagbox{\textbf{Strategies}}{\textbf{Issue Types}}  & \textbf{T1} & \textbf{T2} & \textbf{T3} & \textbf{T4} & \textbf{T5} & \textbf{T6} & \textbf{T7} & \textbf{T8} & \textbf{T9} & \textbf{T10} & \textbf{T11} \\ \hline \hline
		\textbf{S1} & 19    &         & 5        &       &       &       & 37    &       &       &       & 20  \\ \hline
		\textbf{S2} &       & 46      &          &       &       &       &       &       & 17    &       &     \\ \hline
		\textbf{S3} &       &         &          &       &       &       &       &       &       & 12    &     \\ \hline
		\textbf{S4} &       &         &          &       &       & 4     &       & 9     &       &       &     \\ \hline
		\textbf{S5} &       &         &          & 2     & 3     &       &       &       &       &       &     \\ \hline
		\textbf{S6} &       &         &          &       &       &       &       &       &       & 3     &     \\ \bottomrule
		\multicolumn{12}{l}{Each cell reports issue count that are fixed by the corresponding strategy.} \\
		[-0.04cm] 
	\end{tabular}
	\egroup
	\vspace{-3mm}
	\label{t3}
\end{table}

\textbf{Strategy 1:}
As shown in Table~\ref{t3}, nearly half (81/177) of our studied ARP issues can be fixed by \emph{completing the permission check and request process}. 
The strategy is effective for fixing issues of Types 1, 3, 7, and 11.
Figure~\ref{fig_patch1}(a) shows a patch that fixes issue \#650~\cite{issue650} (Type~7) in {\mycode commons-app}.
As we can see, developers added permission check and request operations before invoking the location-sensing APIs, which require dangerous permissions.
Although the strategy is straightforward, in practice, applying it to fix ARP issues is non-trivial due to two reasons.
First, as discussed in issue \#2110~\cite{issue2110} of {\mycode K-9}, since users may revoke granted permissions, developers should make sure that the permission check and request operations are performed at each call site of permission-protected APIs.
Second, from developers' discussions in issue reports (e.g., issue \#76~\cite{issue76} of {\mycode QRCodeReaderView}), we can tell that if the permission check and request operations are performed after the critical task that uses the permission-protected APIs is started, the task may fail once users deny the permission requests. 
Therefore, developers should identify appropriate locations in their source code for adding such operations to avoid the abortion of the critical tasks.

\textbf{Strategy 2:} The second common strategy is to \emph{check SDK versions before requesting permissions or invoking permission-protected APIs}.
Developers adopted this strategy to fix 63 issues of Types 2 and 9, which are essentially compatibility issues induced by platform evolution.
For example, Figure~\ref{fig_patch1}(b) shows the patch for issue \#269~\cite{issue269} (Type 9) in {\mycode device-info}. 
As we can see, the developers revised the code to make it work on API level 26 and above. 

\textbf{Strategy 3:} ARP Issues of Type 10 are caused by libraries' setting of {\mycode maxSdkVersion} in the {\mycode <uses-permission>} elements.
To fix such issues, developers often chose to \emph{revise their app's {\mycode <uses-permission>} settings} to counteract the SDK version constraints set by the libraries.
We observed that 12 issues were fixed in this way.
For example, as shown in Figure~\ref{fig_patch1}(c), when fixing issue \#168~\cite{Apptentive_issue168} in {\mycode Apptentive-android}, developers added the setting {\mycode tools:node= ``replace''} when declaring the permission in the app's {\mycode manifest} file~\cite{manifestmerge}.\footnote{An alternative is to set {\mycode tools:remove=``attribute''}}


\textbf{Strategy 4:} Issues of Types 6 and 8 are essentially API misuse issues.
We found that developers often fix them by \emph{revising the usage of permission check or request APIs}.
In our dataset, 13 issues were fixed in this way.
Figure~\ref{fig_patch1}(d) gives the patch of issue \#2~\cite{issue2} (Type 8) in {\mycode Barcode-Reader}.
The developers originally used the API {\mycode ActivityCompat.requestPermissions()} to request permissions.
However, this API is designed for {\mycode Activity} components instead of {\mycode Fragment}s.
To fix the issue, developers replaced the misused API with another API specifically designed for {\mycode Fragment}s.


\textbf{Strategy 5:}
Issues of Types 4 and 5 can only occur on certain Android device models.
To fix such device-specific issues, developers often \emph{check device information (e.g., the manufacturer) before using permission-protected APIs}.
As an example, Figure~\ref{fig_patch1}(e) gives a patch of issue \#766~\cite{issue766} in {\mycode Katzer}.
The purpose of the patch is to adapt the app to work on Xiaomi devices with legacy platforms, which allow users to revoke permissions.
As we can see, to avoid app crashes caused by invoking APIs whose required permissions are revoked by users, the revised code launches the system settings app to ask users to re-grant the revoked permissions.

\textbf{Strategy 6:} Besides Strategy 3, we found that some developers also fixed issues of Type 10 by \emph{coordinating with library developers to remove the SDK version constraints on dangerous permissions.}
For example, the library {\mycode Hockey-SDK-Android}~\cite{HockeySDK} set {\mycode maxSdkVersion} for the permission {\mycode WRITE\_EXTERNAL\_STORAGE} in its manifest file, which caused ARP issues in three apps~\cite{issue382, issue356, issue373}.
We observed that the app developers requested the library developers to remove the {\mycode maxSdkVersion} setting to avoid its impact on client apps.
The pull request~\cite{PR401} for addressing this issue, which is shown in Figure~\ref{fig_patch1}(f), was merged by the library developers.

\section{Field Survey and Interview}
\label{sec:Industry}
The empirical study in Section~\ref{sec:empirical} focused on data from open-source projects.
To explore RQ4\textendash RQ5 and deepen our understanding of ARP issues, we also performed an online survey and interviews among Android practitioners.
Specifically, to answer \textbf{RQ4}, we designed a questionnaire based on the knowledge gained from the empirical study and invited 250 Android practitioners from seven representative IT companies to participate in the survey.
To answer \textbf{RQ5}, we interviewed six experienced Android practitioners with different backgrounds, which allowed us to further understand ARP issues from different perspectives.
In the following, we present our study design and major findings.

\subsection{Study Design and Setup}

\begin{table*}[]
	\footnotesize
	\centering
	\caption{Basic information of our interviewees}
	\bgroup
	\setlength\tabcolsep{8.0pt}     
	\def\arraystretch{1.1}
	\vspace{-2mm}
	\begin{tabular}{c|c|l|c}
		\toprule
		\rowcolor[HTML]{EFEFEF} 
		\textbf{Practitioner} & \textbf{Company}            & \multicolumn{1}{c|}{\cellcolor[HTML]{EFEFEF}\textbf{Position \& Work Experience}} & \textbf{Product \& Downloads} \\ \hline \hline
		&                             & Android Development Leader                                                        & \emph{Tik Tok}$^a$                       \\ \cline{3-4} 
		\multirow{-2}{*}{P1}  & \multirow{-2}{*}{\emph{ByteDance}} & 6 years Android app development                                                   & 1,000,000,000+ downloads      \\ \hline
		&                             & Android Engineer/Android Tester                                                   & \emph{Tik Tok}$^a$                       \\ \cline{3-4} 
		\multirow{-2}{*}{P2}  & \multirow{-2}{*}{\emph{ByteDance}} & 5 years Android app development \& 2 years Android app testing                    & 1,000,000,000+ downloads      \\ \hline
		&                             & Android Engineer                                                                  & \emph{Meituan dianping}$^b$              \\ \cline{3-4} 
		\multirow{-2}{*}{P3}  & \multirow{-2}{*}{\emph{Meituan}}   & 5 years Android app development                                                   & 1,000,000+ downloads          \\ \hline
		&                             & Android Engineer                                                                  & \emph{Kwai}$^c$                          \\ \cline{3-4} 
		\multirow{-2}{*}{P4}  & \multirow{-2}{*}{\emph{Kuaishou}}  & 6 years Android app development                                                   & 100,000,000+ downloads        \\ \hline
		&                             & Android Engineer                                                                  & \emph{Sogou PinYin}$^d$                  \\ \cline{3-4} 
		\multirow{-2}{*}{P5}  & \multirow{-2}{*}{\emph{Sogou}}     & 8 years Android app development                                                   & 10,000,000+ downloads         \\ \hline
		&                             & Android Engineer/Android Tester                                                   & -                             \\ \cline{3-4} 
		\multirow{-2}{*}{P6}  & \multirow{-2}{*}{A startup} & 4 years Android app development \& Android app testing                            & -                             \\  \bottomrule
		\multicolumn{4}{l}{\emph{\textbf{a}. https://play.google.com/store/apps/details?id=com.zhiliaoapp.musically;}  \emph{\textbf{b}. https://play.google.com/store/apps/details?id=com.sankuai.meituan;}} \\
		\multicolumn{4}{l}{\emph{\textbf{c}. https://play.google.com/store/apps/details?id=kwai.video.community;}  \emph{\textbf{d}. https://play.google.com/store/apps/details?id=com.sohu.inputmethod.sogou}} \\[-0.04cm] 
	\end{tabular}
	\vspace{-4mm}
	\egroup
	\label{table4}
\end{table*}

\subsubsection{Online Survey}
\label{sec:Survey}

To learn practitioners' experiences in dealing with ARP issues, which cannot be learned via analyzing the data of open-source projects, we invited 250 industrial practitioners to participate in an online survey.
These invited participants work for IT companies of different scales, including \emph{Tencent}\footnote{https://www.tencent.com/en-us/}, \emph{ByteDance}\footnote{https://www.bytedance.com/en/}, \emph{Baidu}\footnote{https://home.baidu.com/}, \emph{Meituan}\footnote{https://www.meituan.com/}, \emph{Sogou}\footnote{https://www.sogou.com/}, \emph{Kuaishou}\footnote{https://www.kuaishou.com/}, and a startup company (the name is anonymized for business concerns).
Most of the companies are well-known globally.
In particular, \emph{Tencent} and \emph{ByteDance} are leading IT companies in China, whose representative app products are \emph{WeChat} and \emph{Tik Tok}, which have attracted billions of users.

\vspace{2mm}
\noindent\textbf{\emph{(1) Questionnaire Design}}
\vspace{2mm}

Our questionnaire consists of two parts: (1) questions to collect practitioners' background information and (2) questions to learn practitioners' experiences in handling Android runtime permissions.\footnote{Our questionnaire is at: https://forms.gle/xTaJ7AtpUp5rcSWUA}

The questions in the first part include:
\begin{itemize}
	\item \textbf{Q1:} \emph{How long have you been developing Android applications?}
	
	0-1 year / 1-3 years / 3-5 years / over 5 years;
	
	\item \textbf{Q2:} \emph{During Android app development, have you dealt with runtime permissions?}
	
	Yes / No.
\end{itemize}

Q1\textendash Q2 help to filter respondents (i.e., the survey proceeds only when the respondents have experiences of dealing with runtime permissions) and allow us to analyze the responses of participants with different experience levels.

The questions in the second part include:
\begin{itemize}
	\item \textbf{Q3:} \emph{Do you deal with the cases where users revoke certain permissions?}
	
	Carefully dealing with all such cases / Dealing with some of such cases / Not dealing with such user interaction scenarios;
	
	\item \textbf{Q4:} \emph{How do you determine which permissions are required for an API?}
	
	Checking official Android documentation / Checking Java annotations in the source code of Android SDK / Searching technical forums such as SO / Based on my own development experiences / Other (fillable);
	
	\item \textbf{Q5:} \emph{Are you used to follow the changes in runtime permission mechanisms when a new version of Android OS is released?}
	
	Yes / No;
	
	\item \textbf{Q6:} \emph{Do you consider the customization of the vanilla Android OS made by mobile device manufacturers when handling runtime permissions?} 
	
	Yes / No;
	
	\item \textbf{Q7:} \emph{Do you examine the use of APIs for checking and requesting permissions?} 
	
	Checking all such API uses / Checking some of such API uses / Not checking such API uses; 

	\item \textbf{Q8:} \emph{Do you use third-party libraries (e.g., \textsc{PermissionsDispatcher}\footnote{\textsc{PermissionsDispatcher} is a third-party that provides a simple annotation-based API to handle runtime permissions (https://github.com/permissions-dispatcher/PermissionsDispatcher).}) that help handle runtime permissions, during the development process?}
	
	Yes / No;

	\item \textbf{Q9:} \emph{If your Android app uses third-party libraries, do you realize how the libraries deal with runtime permissions?}
	
	Yes / No;
	
	\item \textbf{Q10:} \emph{Do you realize the impacts on the runtime permission handling logics brought by introducing third-party libraries into your Android app?}
	
	Yes / No;

	\item \textbf{Q11:} \emph{Do you use any tools to help check whether your Android apps handle runtime permissions properly?}
	
	Yes / I Know about such tools but I have never used any of them / I have never used such tools and know little about them;

	\item \textbf{Q12:} \emph{Do you think a tool that helps check whether Android apps properly handle runtime permissions would be useful?}
	
	Very useful / Useful / Maybe useful / Useless.
	
\end{itemize}


\begin{figure}[t!]
	\centering
	\includegraphics[width=0.48\textwidth]{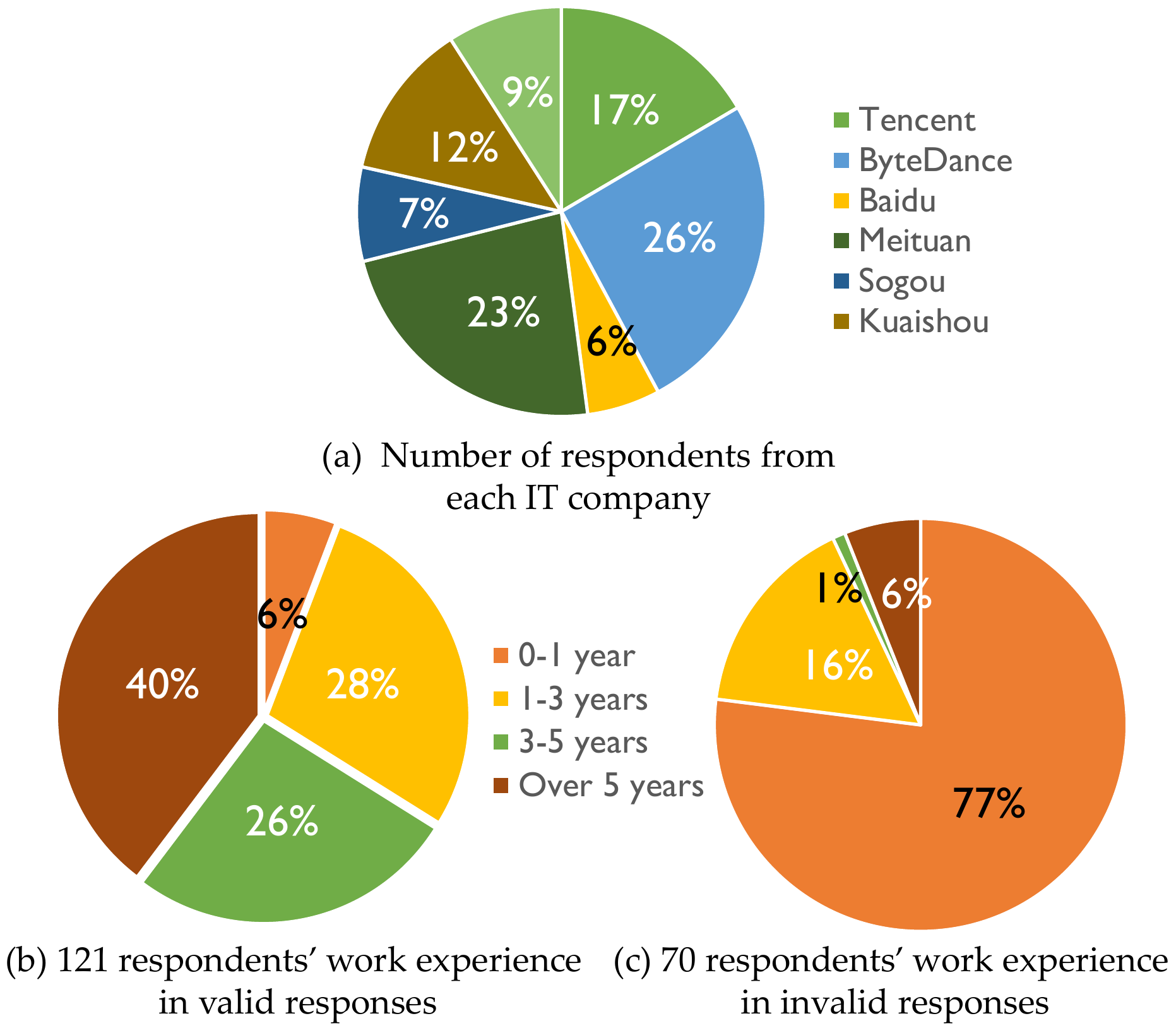}
	\vspace{-2mm}
	\caption{Demographics of participants in our survey}
	\label{Pie3}
	\vspace{-4mm}
\end{figure}

\textbf{Q3-Q10} were formulated based on our empirical findings in Section\ref{sec:empirical}.
Specifically, these questions focus on the factors that can induce ARP issues.
Combining the responses from participants with the root causes of different ARP issues summarized in the previous study, we can effectily learn real practices that may cause or prevent ARP issues, which may provide insights for future research on ARP issue diagnosis.
We additionally designed \textbf{Q11} and \textbf{Q12} to learn practitioners' opinions and requirements of tools to combat ARP issues.
In the end of the questionare, we also asked the participants to leave their contact information if they were available for an interview to further discuss the problems of Android runtime permissions.

\vspace{2mm}
\noindent\textbf{\emph{(2) Survey Participants}}
\vspace{2mm}

In total, we received 191 responses from practitioners working in seven IT companies.
Figure~\ref{Pie3}(a) presents the demographic information of the participants.
121 out of 191 respondents (63.4\%) have experiences of dealing with runtime permissions and their responses are considered valid.
Figure~\ref{Pie3}(b) and (c) show the experiences of the respondents providing valid and invalid responses, respectively.
Among the 70 respondents without prior experiences of handling runtime permissions, 77\% have 0-1 year of Android app development experience.
Comparatively, most of the 121 respondents with experiences of handling runtime permissions are experienced Android app developers (3\textendash 5 years).
As such, further analysis of survey results was performed on the 121 valid responses.
We will present our findings in Section~\ref{sec:RQ4}.

\subsubsection{Practitioner Interviews}
\label{sec:Interviews}

To understand the common challenges faced by practitioners when they encounter ARP issues, we further conducted audio interviews with the survey respondents.

\vspace{2mm}
\noindent\textbf{\emph{(1) Interviewees}}
\vspace{2mm}

Fifteen participants of our online survey left their contact information and volunteered to participate in our interviews.
Since interviews are time-consuming, we invited six of these volunteers for the audio interviews.
Table~\ref{table4} provides the information of the six interviewees.
As we can see from the table, our interviewees are experienced in Android app development: all of them have at least four years of experience.
They also have different positions in companies of different scales.
P1 is an Android development leader working on \emph{Tik Tok} (from \emph{ByteDance}, a leading IT company in China), which is one of the most popular short video social app all over the world.
P2 is P1's colleague, who is a senior Android engineer having fives years of development experience and two years of testing experience.
In particular, P1 and P2 are responsible for runtime permissions-related matters; 
P3, P4 and P5 are senior Android engineers from \emph{Meituan}, \emph{Kuaishou}, and \emph{Sogou}, respectively, and participated in the development of globally well-known app products \emph{Meituan dianping}, \emph{Kwai} and \emph{Sogou PinYin}, respectively.
P6 is responsible for both app development and testing in a small startup company.

\vspace{2mm}
\noindent\textbf{\emph{(2) Interview Design}}
\vspace{2mm}

We conducted semi-structured interviews with the six practitioners.
The interviews have three parts and contain a general question designed based on RQ5 and a series of open-ended questions.
Specifically, our question in the first part was phrased as  \emph{``What challenges have you encountered when dealing with ARP issues and what solutions have you adopted to solve the problems?''}
The aim of this question was to initiate the discussion, allowing the interviewees to freely share their experience without focusing on any specific type of ARP issues.
In the second part of the interview, we improvised questions based on the interviewees' answers to our first question.
For instance, we asked the scenarios or root causes of their encountered ARP issues and then, if they provided clear description, we proceeded with more detailed questions to further understand the related challenges.
In the third part, we asked the interviewees' requirements and expected features for a tool that can help them deal with ARP issues, so as to inspire the future research.

The length of interviews varied between 24 and 43 minutes, with an average of 32 minutes.
We sought the interviewees' consent to make audio recording of all the interviews for subsequent analysis.\footnote{The recordings are available at http://arp-issues.github.io/.}
We will discuss our findings in Section~\ref{sec:RQ5}.
\begin{figure*}[t!]
	\centering
	\includegraphics[width=0.9\textwidth]{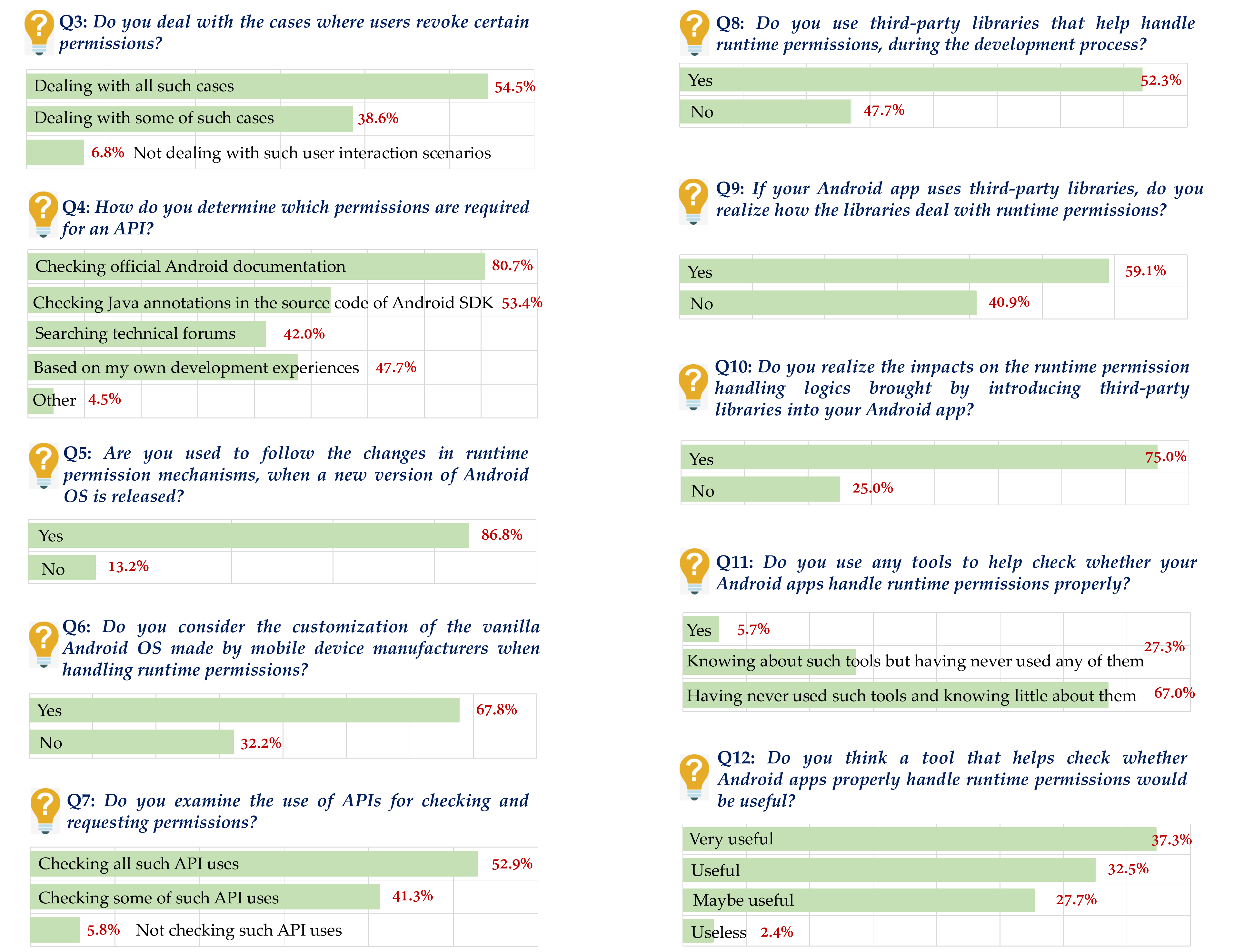}
	\vspace{-2mm}
	\caption{Practitioners' responses to our questionnaire} 
	\label{fig_Survey}
	\vspace{-2mm}
\end{figure*}

\subsection{RQ4: Industrial Practices}
\label{sec:RQ4}

Figure~\ref{fig_Survey} gives the distribution of practitioners' responses to our questionnaire.
As previously described, questions Q3\textendash Q10 were formulated based on the root causes of our observed ARP issues: the scenario mentioned in Q3 may cause Type 1 issues; Q4 and Q5 are related to issues of Types 2 and 3;
Q6 mentions the cause of issues of Types 4, 5 and 6;
Q7 and Q8 are related to issues of Types 7 and 8;
the factors described in Q9 and Q10 can induce issues of Types 9, 10 and 11.
The purpose of asking them is to learn industrial practices on permission handling, which may cause or help prevent ARP issues.
Now we analyze the practitioners' responses and present our findings.

\vspace{6pt}
\noindent\begin{tcolorbox}
	\small
	\noindent \textbf{Finding 11:} \emph{A large percentage of real-world practitioners have not considered all situations where users revoke granted permissions. 
	}
\end{tcolorbox}
\vspace{2pt}

From the responses to Q3, 38.6\% practitioners deal with permission revocation cases partially, and 6.8\% practitioners are not aware of such abnormal user interactions. 
Such omission in handling permissions can easily cause runtime exceptions, especially when legacy apps run on new platforms, as we have discussed in Section~\ref{sec:empirical}. 

\vspace{6pt}
\noindent\begin{tcolorbox}
	\small
	\noindent \textbf{\textit{Finding 12:}} \emph{Practitioners have different ways of determining API-DP mappings for handling runtime permissions. Many of them also pay attention to the evolution of the runtime permission mechanism.
	}
\end{tcolorbox}
\vspace{2pt}

For Q4 and Q5, we found that practitioners may determine API-DP mappings in many ways, including checking the official Android documentation (80.7\%), checking Java annotations in the source code of Android SDK (53.4\%), searching technical forums such as SO (42\%), and relying on their own development experiences (47.7\%).
86.8\% of respondents concern the changes in the runtime permission mechanism when a new version of Android OS is released. 
These results suggest that many practitioners have realized that the evolution of API-DP mappings and the runtime permission mechanism can induce ARP issues.
However, thoroughly checking such evolution and validating its impact on dealing with runtime permissions are non-trivial tasks for practitioners.
We have witnessed many real ARP issues of Types 2 and 3 in open-source projects (see Section~\ref{sec:Pattern A}).

\vspace{6pt}
\noindent\begin{tcolorbox}
	\small
	\noindent \textbf{Finding 13:} \emph{Many practitioners do not realize the impact of mobile device manufacturers' customization and third-party libraries' configurations on Android apps' handling of runtime permissions. 
	}
\end{tcolorbox}
\vspace{2pt}

From the answers to Q6, we can see that although many Android device manufacturers make various permission-related system customization, which can easily induce ARP issues of Types 4, 5 and 6, 32.2\% of the surveyed practitioners do not realize such customization on specific device models. 
Similarly, as we can see from the answers to Q9 and Q10, although many practitioners might have realized that ARP issues can be caused by inconsistent permission handling between app modules and third-party libraries (issues of Types 9, 10 and 11), there are still 40.9\% respondents not realizing how third-party libraries deal with runtime permissions, and there are 25\% respondents paying no attention to the impacts on runtime permission handling logics brought by third-party libraries.

\vspace{6pt}
\noindent\begin{tcolorbox}
	\small
	\noindent \textbf{Finding 14:} \emph{Tools that can help manage runtime permissions or detect improper handling of runtime permissions can be very useful to practitioners.
	}
\end{tcolorbox}
\vspace{2pt}

The answers to Q7 and Q8 show that only 52.9\% practitioners carefully examine the uses of APIs for checking and requesting permissions, possibly because the process is troublesome. 
The missing examination of such API uses may induce ARP issues of Type 7 (i.e., incomplete permission check and request process) and 8 (misusing permission request APIs).
To avoid making mistakes in runtime permission handling, 52.3\% of our surveyed practitioners adopt third-party tools to help deal with runtime permissions.
In addition to tools that help manage runtime permissions, the majority (69.8)\% of respondents are also interested in tools that can help detect ARP issues, as we can tell from their answers to Q12.
Two survey participants even left comments, saying that their companies have developed such tools for internal use.
However, according to the answers to Q11, 67\% of the practitioners know little about such tools.
These results show that there is a real need of automated techniques to help developers effectively manage runtime permissions and combat ARP issues.
Future research may focus on developing such techniques and make them available to real-world Android practitioners.

\subsection{RQ5: Challenges Faced by Practitioners}
\label{sec:RQ5}

\vspace{6pt}
\noindent\begin{tcolorbox}
	\small
	\noindent \textbf{Finding 15:} \emph{Due to the differences in production functionalities, business requirements, and testing schemes, industrial practitioners faced different challenges when dealing with runtime permissions.
	}
\end{tcolorbox}

In our interviews, the participants shared with us the problems they met when dealing with Android runtime permissions and how they solved the problems.
In the following, we discuss their viewpoints from four aspects:

\begin{itemize}
	
	\item \textbf{Android platform evolution.} P2, P3, and P5 stated that they have encountered ARP issues introduced by the evolution of API-DP mappings and the permission group mechanism.
	Both P3 and P5 said that they would pay special attention to permission-related information in the release notes of Android SDK, especially for the cases of ``permission contractions''.
	P3 shared a ``permission contraction'' case: ``At API level 28, one of their APIs required the {\mycode READ\_PHONE\_STATE} permission, which can be requested at runtime from users.
	During the evolution, this API turned to require the {\mycode READ\_PRIVILEGE\_PHONE\_STATE} permission.
	However, {\mycode READ\_PRIVILEGE\_PHONE\_STATE} is a system permission and could not be granted to apps since API level 29.''
	As such, to solve the ARP issue, they adapted the permission request on different versions of Android platforms.
	P2 emphasized: ``\emph{We would not upgrade the apps' {\mycode targetSdkVersion} until the new features of upgraded Android SDK version have been adequately tested by our testing group, to avoid ARP and other incompatibility issues.}''
	On the other hand, P1, P4, and P6 said that they rarely encountered such types of issues, since their products only involve few common dangerous permissions, such as {\mycode Camera}, {\mycode Location}, {\mycode Storage}, etc., and in their apps, the dangerous permissions are required by a set of stable APIs.	
	
	\item \textbf{Device manufacturers' customization.} P1, P3, P4, and P5 complained that they have seen lots of ARP issues caused by customized platforms.
	They encountered two main challenges when dealing with such issues:
	(1) before the runtime permission model was introduced into \emph{vanilla} Android OS, several device manufacturers (e.g., Huawei and Xiaomi) had already implemented their own runtime permission mechanism into mainstream products;
	(2) different device models customized the runtime permission mechanism in different ways, and it is difficult to deal with the varieties in Android apps.
	As a development leader in an leading IT company, P1 stated ``\emph{Our testing group would verify apps' permission requests on various mainstream devices to identify the affects induced by their customizations. Development group would then add checking statements for device models and perform the adaptions accordingly.}''
	These experiences are inline with our empirical findings via investigating data from the open-source community. The expert's sharing can also provide implications for future research to develop automated techniques to combat ARP issues of Types 4, 5, and 6.
	
	\item \textbf{Improper permission check and request.} Both P2 and P6 considered improper permission check and request as their major challenges when dealing with runtime permissions.
	They stated that in the diversified and complicated permission handling scenarios, the cases of missing check and request steps for runtime permissions frequently occurred.
	In particular, P2 stated ``\emph{To avoid such negligence, we should clearly know which Android APIs are protected by permissions and carefully program our apps to request permissions at runtime and handle various user actions.}''
	P6 stated that as their products become mature, they adopted third-party tools such as \textsc{PermissionsDispatcher} to simplify the permission check and request process.
	These tools alleviated their burden of writing a bunch of check and request statements and are helpful in keeping the source code clean and safe.
	As such, in P6's company, there are fewer ARP issues in recent two years.
	
	\item \textbf{Interference by third-party libraries.} From the feedback of interviewees, we found that the degree of the interference by third-party libraries in terms of ARP issues may be highly correlated with the adopted open-source libraries and the manpower companies invested in app testing.
	P1 and P2 stated that they rarely encountered library-interfered issues, since most of their integrated libraries are closed-source, which are customized for their apps and have been thoroughly tested.
	P3, P4, and P5 said that they occasionally experienced ARP issues introduced by third-party libraries.
	In their companies, only a subset of libraries are adopted from open-source community.
	P3 emphasized ``\emph{We should pay special attention to the permission configurations in third-party libraries to prevent our apps from inheriting improper configurations.}''
	Comparatively, P6 stated that they often encountered such issues, because in their company, which is a small startup, most of the app modules are developed based on open-source libraries.
	As a good practice, in the entry methods that directly invoke the third-party library APIs, they encapsulated the permission processing operations, to ease the code evolution or library substitution in the future.
	In his view, library developers should \textit{clearly document the need of requesting dangerous permissions in client apps}, if the request should not be performed in the layer of libraries.
	Besides, he also appealed to library developers to \textit{refrain from setting SDK version constraints on dangerous permissions} to avoid affecting client apps.
	P6's ideas can help proactively prevent ARP issues of  Types 9, 10, and 11.
	
\end{itemize}

\vspace{6pt}
\noindent\begin{tcolorbox}
	\small
	\noindent \textbf{Finding 16:} \emph{Practitioners have various requirements for tools that can detect ARP issues.
	}
\end{tcolorbox}

Interviewees' experiences are complementary to our empirical findings observed via analyzing the data from the open-source community.
Besides the experience sharing, they also expressed their requirements for tools that can help detect ARP issues:
(1) the tool should construct a testing framework to analyze how their apps' permission requests behave on different device models and Android versions (from P1, P2, P3, P4, and P5); 
(2) the tool should simulate various permission-control behaviors of users (e.g., denying the permission requests) and analyze how the apps would behave in such scenarios (from P2 and P6);
(3) the tool should identify all the permission-protected APIs, especially in the third-party libraries and analyze whether the permission check and request operations have been properly performed (from P2 and P6);
(4) the tool should analyze libraries' permission configurations and validate whether their permission check and request operations behave properly after being integrated into the apps (from P6).
We believe that the above analyses of the interview results along with our issue taxonomy can guide future research to design automated tools to help Android developers combat ARP issues.

\section{RQ6: Performance of Existing ARP Issue Detectors}
\label{sec:Implications for Future Research}


As we mentioned ealier, in recent years, there have been several tools designed to detect ARP issues in Android apps.
With the 11 types of common ARP issues identified in our previous study, we are intrigued to know how these tools would perform in practice.
Toward this end, we built a benchmark consisting of 94 real ARP issues and used it to study the performance of three publicly accessible ARP issue detectors.
We report this study in the following.

\subsection{Study Design and Setup}

\begin{table*}[]
	\footnotesize
	\label{T5}
	\centering
	\caption{The performance of existing ARP issue detectors}
	\bgroup
	\setlength\tabcolsep{3.5pt}     
	\def\arraystretch{1.2}
	\begin{tabular}{c|c|c|c|c|c|c|c|c|c|c}
		\toprule
		\rowcolor[HTML]{EFEFEF}\textbf{Tools}  
		& \multicolumn{1}{c|}{\cellcolor[HTML]{EFEFEF}\textbf{Type 1}} & \multicolumn{1}{c|}{\cellcolor[HTML]{EFEFEF}\textbf{Type 2}} & \multicolumn{1}{c|}{\cellcolor[HTML]{EFEFEF}\textbf{Type 3}} &  \multicolumn{1}{c|}{\cellcolor[HTML]{EFEFEF}\textbf{Type 7}} & \multicolumn{1}{c|}{\cellcolor[HTML]{EFEFEF}\textbf{Type 8}} & \multicolumn{1}{c|}{\cellcolor[HTML]{EFEFEF}\textbf{Type 9}} & \multicolumn{1}{c|}{\cellcolor[HTML]{EFEFEF}\textbf{Type 10}} & \multicolumn{1}{c|}{\cellcolor[HTML]{EFEFEF}\textbf{Type 11}} & \multicolumn{1}{c|}{\cellcolor[HTML]{EFEFEF}\begin{tabular}[c]{@{}c@{}}\textbf{\# Detected issues}\\ \textbf{(Issue Detection Rate)}\end{tabular}} & \multicolumn{1}{c}{\cellcolor[HTML]{EFEFEF}\begin{tabular}[c]{@{}c@{}}\textbf{\# False alarms}\\ \textbf{(False Alarm Rate)}\end{tabular}} \\ \hline
		\cellcolor[HTML]{EFEFEF}\textsc{Lint}     & 0/9   & 0/22  & 2/5  & 11/18 & 0/4  & 0/11  & 0/10   & 7/15 & 20 (21.3\% = 20/94)                 & 11 (55.0\% = 11/20)              \\ \hline
		\cellcolor[HTML]{EFEFEF}\textsc{ARPDroid} & 6/9 & 0/22  & 1/5  & 8/18  & 0/4  & 0/11  & 0/10   & -    & 15 (19.0\% = 15/79)                 & 15 (100.0\% = 15/15)              \\ \hline
		\cellcolor[HTML]{EFEFEF}\textsc{RevDroid} & 6/9 & 0/22  & 2/5  & 9/18  & 0/4  & 0/11  & 0/10   & -    & 17 (21.5\% = 17/79)                 & 17 (100.0\% = 17/17)              \\ \bottomrule
	\end{tabular}
	\egroup
	\label{tab:tool_perf}
\end{table*}

\textbf{ARP issue detectors.} To address RQ6, we selected the latest version of three existing tools for empirical evaluation and comparison:

\begin{itemize}
	
	\item \textsc{Lint}~\cite{Lint}, a static analyzer that scans the source files of Android projects to find common bugs, including ARP issues. 
	As it is integrated in Android Studio, Lint has been widely-used by developers~\cite{wei2017oasis}.
	
	\item \textsc{ARPDroid}~\cite{dilhara2018automated}, a technique to detect and fix compatibility issues related to runtime permissions by analyzing the {\mycode apk} file of an Android app.
	
	\item \textsc{revDroid}~\cite{fang2016revdroid}, a tool to identify ARP issues that arise when users revoke permissions by analyzing the {\mycode apk} file or an Android app.
	
\end{itemize}

\begin{figure}[t!]
	\centering
	\includegraphics[width=0.5\textwidth]{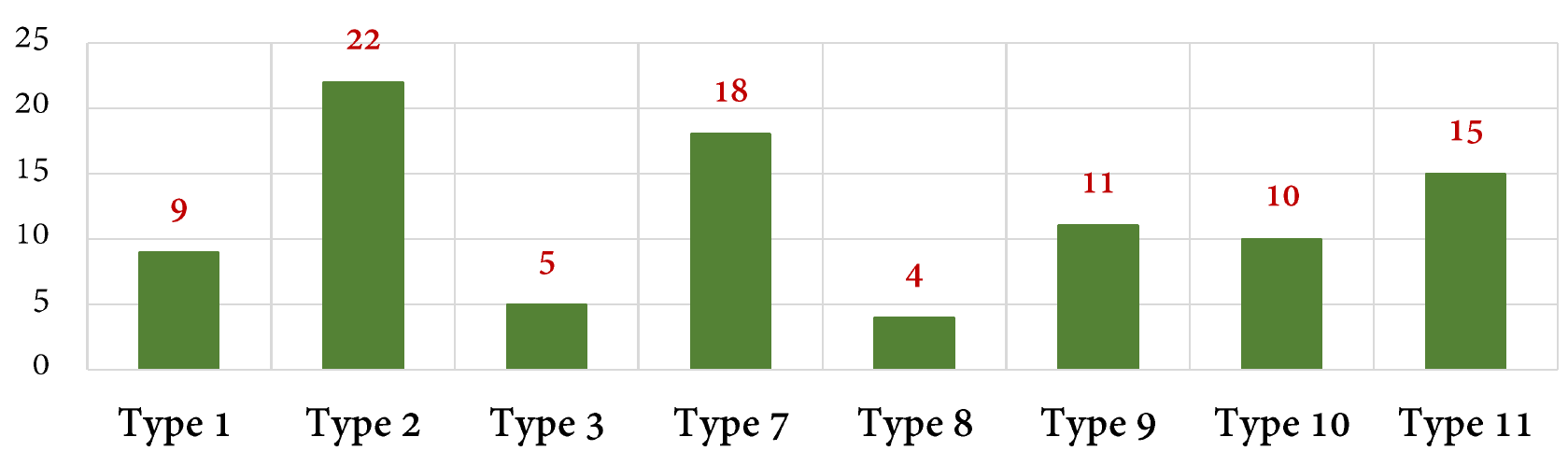}
	\vspace{-4mm}
	\caption{Distribution of each type of selected ARP issues in the \textsc{ARPBench}}
	\label{Benchmark}
	\vspace{-6mm}
\end{figure}

\textbf{Benchmark.} For the experiments, we also need a benchmark of ARP issues.
As there is no readily available benchmark, we selected a subset of the real ARP issues collected in our empirical study (Section~\ref{sec:RQ2}) and constructed a benchmark, named \textsc{ARPBench}, by ourselves.
We did not use all ARP issues from our empirical study dataset due to three major reasons.
First, for each ARP issue, we need both buggy and patched app versions to evaluate the three tools' performance.
Some issues investigated in our empirical study have not been fixed by developers yet and hence they are not suitable for the tool evaluation.
Second, compiling open-source Android apps is a cumbersome process.
Many open-source apps rely on specific libraries, which can be hard to find. 
What is more, open-source apps do not always have build scripts or clear compilation instructions.
Third, for the ARP issues caused by manufacturers' customization, we could not figure out the exact device models on which the issues would occur.
Besides, to the best of our knowledge, there are no existing detectors for such device-specific issues.
Therefore, we ignored these issues (Types 4\textendash6).
With the above constraints and considerations, we finally selected 94 ARP issues from the 199 issues collected in Section~\ref{sec:RQ2} to construct \textsc{ARPBench}.
Figure~\ref{Benchmark} shows the distribution of the ARP issues in \textsc{ARPBench}.
It is worth noting that all of our collected Type 11 issues (i.e., library-interfered ARP issues) are reported in the third-party libraries' issue trackers.
We could not find the corresponding apps that suffered from the ARP issues and therefore could not prepare {\mycode apk} files to evaluate \textsc{ARPDroid} and \textsc{revDroid}.
The 15 issues of Type 11 were only used to evaluate \textsc{Lint}, which directly analyzes source code for issue detection.

\textbf{Metrics.} Following a recent study that evaluates resource leak detectors for Android apps~\cite{droidleaks}, we evaluated the \textit{issue detection rate} and \textit{false alarm rate} of the three detectors, which are defined in Equations (1) and (2), respectively.
Specifically, we leveraged the buggy app versions to evaluate the issue detection rate of a detector $t$, denoted as $IDR(t)$.
For each selected issue in our benchmark, if the detector $t$ reports a warning that describes an ARP issue at the concerned issue location of the buggy app version, we consider this warning as a true one, i.e., the detector $t$ successfully detects the ARP issue.
Otherwise, we consider that the tool misses the ARP issue.
After evaluating $t$ with all applicable ARP issues, $t$'s issue detection rate can be calculated by Equation~(1).
Furthermore, we leveraged the patched app versions to evaluate the false alarm rate of each detector $t$,  denoted as $FAR(t)$.
For each issue that is detected by the detector $t$, if $t$ still reports a warning after analyzing the patched app version, we consider this warning as a false alarm.
Similarly, after evaluating $t$ with all applicable patched versions,  $t$'s false alarm rate can be calculated by Equation~(2).
Note that during the experiments, we would ignore the warnings that are not related to our selected ARP issues, due to the lack of reliable ground truth to judge whether the warnings are true ones or not.

\begin{equation}\label{eq1}
\footnotesize
IDR(t) = \frac{\#\text{ARP issues detected by \emph{t} on buggy app versions}}{\#\text{Buggy ARP issues experimented on \emph{t}}}
\end{equation}

\begin{equation}\label{eq2}
\footnotesize
FAR(t) = \frac{\#\text{false alarms reported by \emph{t} on patched app versions}}{\#\text{Patched ARP issues experimented on \emph{t}}} 
\end{equation}

\subsection{Experiment Results}
Table~\ref{tab:tool_perf} reports the experiment results of the three ARP issue detectors. We discuss the main findings below.
\vspace{2pt}
\noindent\begin{tcolorbox}
	\small
	\noindent \textbf{Finding 17:} \emph{The three detectors are weak in detecting the 11 common types of ARP issues. In particular, none of the detectors can detect ARP issues of Types 2, 4\textendash6, and 8\textendash10, which are caused by the evolution of API-DP mappings, device manufacturers' customization, and interference of third-party libraries.}
\end{tcolorbox}
\vspace{2pt}

We observed that the tools' poor performance in terms of issue detection rate is due to their following limitations:

\begin{itemize}
	
	\item \textit{Not aware of the effects caused by Android platform evolution}. 
	The three detectors do not take the evolution of API-DP mappings across different Android platforms into consideration when detecting ARP issues.
	As such, they missed the issues of Types 2 and 9.
	
	\item \textit{Not aware of the effects caused by conflicting configurations between apps and their referenced libraries}. 
	None of the three evaluated detectors could identify the conflicting configurations between apps and their referenced libraries on runtime permissions and analyze the effects of such conflicts. Due to this reason, the detectors missed Type 10 issues.
	
	\item \textit{Not aware of the misuse of permission request APIs}.
	\textsc{Lint} and \textsc{revDroid} only check the uses of permission check APIs and ignore permission request APIs.
	Even worse, \textsc{ARPDroid} performs coarse-grained analysis and do not check the correctness of the uses of both permission check and request APIs.
	As a result, they do not support detecting Type 8 issues.
	
	\item \textit{Not aware of the effects caused by platform customizations}. 
	Despite the lack of Types 4\textendash6 issues for experiments, to the best of our knowledge, the three detectors are not capable of detecting the ARP issues caused by various customizations made by device manufacturers.
	
	\item \textit{No support for legacy apps}. Lastly, by further analyzing the \textsc{ARPBench} and \textsc{Lint}'s source code, we found that \textsc{Lint} did not check ARP issues of Type 1 for legacy apps. This is also a limitation of \textsc{Lint}.
	
\end{itemize}

\vspace{6pt}
\noindent\begin{tcolorbox}
	\small
	\noindent \textbf{Finding 18:} \emph{Although existing detectors can detect issues caused by incomplete permission check and request (Types 1, 3, 7, and 11), they could still miss many real issues in \textsc{ARPBench}.}
\end{tcolorbox}
\vspace{2pt}

The three detectors missed some issues of Types 1, 3, 7, and 11 in \textsc{ARPBench} because of the following two reasons:

\begin{itemize}
	
	\item \textit{Outdated or incomplete API-DP mappings}. \textsc{ARPDroid} and \textsc{revDroid} detect ARP issues based on the API-DP mappings obtained by \textsc{PScout}~\cite{au2012pscout} (i.e., a study on permission specifications).
	However, \textsc{PScout} only produced the permission specifications for Android 4.1.1\textendash5.1.1, which are outdated.
	Besides, as mentioned in Section~\ref{sec:Distribution}, since Android 6.0 (API level 23), Google started to document permission specifications in two ways: 1) using Java annotation {\mycode @requiresPermission} to associate APIs with permissions and 2) using {\mycode @link android.Manifest.permission\#} to describe an API's required permissions.
	By analyzing the source code of \textsc{Lint}, we found that it collects the API-DP mappings only by analyzing the Java annotations in the source code of Android SDK.
	As such, its API-DP mappings are not complete.
	
	\item \textit{Not sensitive to all the steps of permission check and request process}. \textsc{ARPDroid} does not support detecting issues caused by the lack of \textit{Step 4}.
	Besides, according to the research paper~\cite{fang2016revdroid} and our analysis of \textsc{Lint}'s source code, \textsc{revDroid} and \textsc{Lint} can only identify ARP issues caused by the lack of \textit{Step 2} during the permission check and request process.
	As a result, the three detectors missed some of Types 1, 3, 7, and 11 issues.
	
\end{itemize}

\vspace{6pt}
\noindent\begin{tcolorbox}
	\small
	\noindent \textbf{Finding 19:} \emph{For the ARP issues caused by incomplete permission check and request or misusing permission request APIs (Types 1, 3, 7, and 11), the three detectors all report a large number of false alarms. The false alarm rate can be as high as 100.0\%.}
\end{tcolorbox}
\vspace{2pt}

\begin{figure}[t!]
	\centering
	\includegraphics[width=0.5\textwidth]{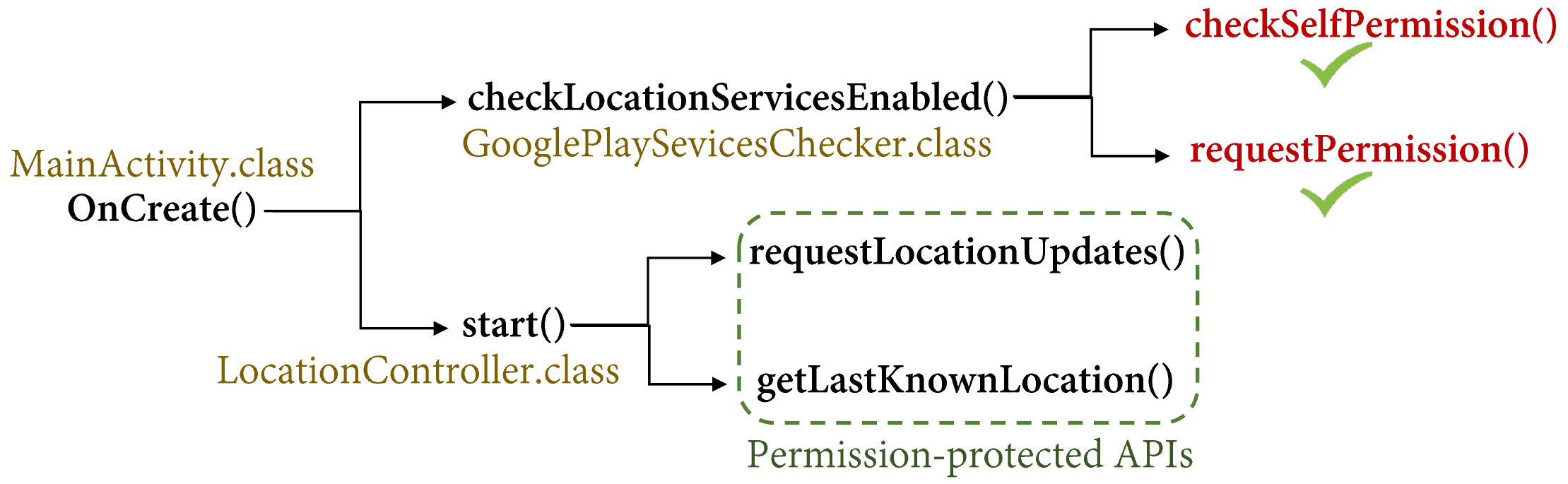}
	\vspace{-4mm}
	\caption{A call chain from the patched app version corresponding to issue \#90 of {\mycode stardroid} to illustrate \textsc{Lint}'s limitation}
	\label{Example case}
	\vspace{-6mm}
\end{figure}

For the 20 ARP issues captured by \textsc{Lint}, we applied \textsc{Lint} to analyze the patched versions and found that
\textsc{Lint} still reported 11 warnings, which are false alarms.
The reason for such a high false alarm rate (55.0\%) is that \textsc{Lint} performs intra-procedural analysis~\cite{7442579} and can only identify the lack of permission check APIs within the method that calls permission-protected APIs.
If the calls of permission check APIs are located in other methods on the callchain, \textsc{Lint} will generate false alarms.
For instance, Figure~\ref{Example case} shows a case extracted from the patched app version corresponding to issue \#90 of {\mycode stardroid}~\cite{issue90}.
In this case, \textsc{Lint} checks whether there is a permission check API in the method {\mycode LocationController.start()}, where the two permission-protected APIs are called.
The developers have correctly inserted the calls to permission check and request APIs into the method {\mycode checkLocationServicesEnabled()} in class {\mycode GooglePlayServicesChecker}, which would be executed before {\mycode LocationController.start()}.
However, since \textsc{Lint} does not perform inter-procedual analysis on such a call chain, it cannot infer the existence of permission check and request steps, and would then report a false alarm.

For the 15 issues reported by \textsc{ARPDroid}, we also applied the tool to analyze the patch versions.
Suprisingly, \textsc{ARPDroid} still identified the lack of permission check and request steps, and inserted the correponding API calls at the call sites of permission-protected APIs, regardless of the existence of such API calls in the source code of the patched verions.
The false alarm rate is 100.0\%.

Similar to \textsc{ARPDroid}, \textsc{revDroid} still reported warnings for all the 17 patched versions.
Although \textsc{revDroid} carefully checks whether there are permission check APIs on the call chains that reach the call sites of permission-protected APIs, it only check the existence of calls to the following APIs: {\mycode checkPermission()}, {\mycode checkCallingOrSelfPermission()}, {\mycode checkCallingPermission()}, and {\mycode checkUidPermission()}.
Since Android 6.0, these APIs have been replaced by {\mycode checkSelfPermission()}.
Since \textsc{revDroid} is not aware of the permission check operations performed using {\mycode checkSelfPermission()}, it reported many false alarms.

\textbf{Suggestions for future ARP issue detector designers}. The existing detectors have very limited support for detecting real-world ARP issues.
An effective ARP issue detection technique should address the following four challenges:
\begin{itemize}

	\item It should be able to deal with diverse types of APR issues with different root causes.
	
	\item It should be powered by complete and precise API-DP mappings. The evolution of Android platform and the customizations performed by device manufacturers should also be taken into consideration.
	
	\item It should be able to analyze the conflicts of configuration attributes in {\mycode manifest} files between an app and its referenced libraries to detect library-interfered issues. 
	
	\item It should be able to identify the permission-protected APIs and their invocation paths in both app modules and libraries.
	In particular, the path identification should be performed both intra-procedurally and inter-procedurally to avoid false alarms or false negatives.
	
\end{itemize}

\section{Threats to Validity}

The validity of our study results may be subject to the following threats.

\textbf{ARP issue collection.} Our collected ARP issues may not be representative and comprehensive.
To mitigate this threat, we followed a rigorous process to collect issues.
First, we studied all SO posts tagged with ``runtime permission'' or containing the general phrase ``Android runtime permission'' to study all relevant issues discussed on SO.
Based on the study results, we then formulated a set of keywords (Table~\ref{t2}) to search for ARP issues in open-source Android projects.
The 415 open-source projects were carefully selected to ensure their diversity and impact (we considered the downloads of the apps, the projects' star and fork counts on GitHub).
In the end, we collected 199 real ARP issues.
We believe that this dataset is of good quality.
We release it for public scrutiny. 

\textbf{Trend analysis.} The number of issues used in the evolution trend analysis (Section~\ref{sec:Evolutionary}) is limited.
To avoid making observations that have little statistical significance, we have already excluded four types of issues with few instances.
Future studies can further validate our findings in this trend analysis, when there are larger-scale datasets.

\textbf{API-DP mappings.} We mined API-DP mappings from the Javadoc of Android SDK for API levels 23\textendash29 (as mentioned earlier, we had to do this by ourselves because existing datasets are outdated).
If the official documentation is not complete, we will miss some mappings.
In future, we will further improve the mining approach and explore more complete API-DP mappings.

\textbf{Errors in manual inspection.} Our study involved much manual work (e.g., analyzing SO posts and issue reports). 
We understand that such a manual process is subject to errors. 
To reduce the threat, we have cross-validated all results for consistency.


\textbf{Incompleteness of the issue taxonomy.} In our study, the taxonomy of ARP issues is obtained by analyzing the collected posts on \textit{SO}.
Although \textit{SO} is one of the most popular Q\&A sites for developers, it may not contain the discussions of all possible types of ARP issues.
Despite that we have verified that our taxonomy could covered all the real ARP issues collected from GitHub when investigating RQ2, future studies may further improve the taxonomy. 


\vspace{-2mm}
\section{Related work}
\label{sec:Related work}
Existing studies~\cite{au2012pscout, fang2014permission, fragkaki2012modeling, nauman2010apex, wei2012permission, wijesekera2015android, li2015iccta, li2014know, bao2016permissions, liu2019automatic, nguyen2017android, bartel2012automatically, bagheri2015detection, bagheri2018formal, bagheri2015covert, liu2014reconciling, felt2012ask, bugiel2013flexible, heuser2014asm, zhou2011taming, agarwal2013protectmyprivacy, olejnik2017smarper, andriotis2016permissions, liu2018large} have investigated many aspects of Android permission system.
We discuss those that are most relevant to our work below.

\textbf{Migration to the runtime permission model.}
Since the release of Android 6.0, researchers have proposed several techniques to help legacy apps automatically migrate to the runtime permission model.
For instance, Dilhara et al.~\cite{dilhara2018automated} presented \textsc{ARPDroid}, a technique to check and enforce the app's conformance to the runtime permission use protocol based on control flow analysis.
Bogd$\check{a}$nas~\cite{bogdanas2017dperm} presented \textsc{DPERM}, a static analysis technique to recommend code locations to insert permission requests into Android apps.
Fang et al.~\cite{fang2016revdroid} designed a tool~\textsc{revDroid}, to identify the issues that arise when permission checks are far away from where the permissions are used. 
These issues can lead to crashes when permissions are revoked by users while permission-protected functions are executing.
However, the above techniques only focus on specific types of APR issues.
For a comprehensive understanding of ARP issues, our work systematically investigates the types, root causes, pervasiveness and seriousness, as well as fixing strategies of the issues related to runtime permissions.


\textbf{User interactions with runtime permission model.}
Researchers have proposed different extensions to runtime permission models to provide users with more control and better management~\cite{liu2014reconciling, felt2012ask, bugiel2013flexible, heuser2014asm, zhou2011taming, agarwal2013protectmyprivacy, olejnik2017smarper}.
Andriotis et al.~\cite{andriotis2016permissions} presented the first study that analyzed Android users' adaptation to the fine-grained runtime permission model, regarding their security and privacy controls.
Considering the accessibility of particular apps to resources, the authors highlighted the persistence of users to allow access to permissions that are directly related to their main functionalities.
To improve the users' understanding of the purposes of requesting permissions, Liu et al.~\cite{liu2018large} conducted a large-scale study on five aspects of runtime-permission rationale messages.
They found that less than one-fourth of apps provided such rationales, which indicated that developers might need further guidance on how to explain the purposes of requesting permissions. 
Shen et al.~\cite{263788} first investigated to what extent current system-provided information could help users understand the scope of permissions and their potential risks.
They took a mixed-methods approach by collecting real permission settings from 4,636 Android users, an interview study of 20 participants, and large-scale Internet surveys of 1559 users.
Their study identified several common misunderstandings on the runtime permission model among users.
Our work considered such user interactions in adverse conditions as an important type of triggering conditions of ARP issues, which provided implications for designing an effective testing technique.

\textbf{Inferring API-permission mappings.}
Several approaches~\cite{au2012pscout, bartel2012automatically, bartel2014static, nguyen2017android, backes2016demystifying, bao2017automated, dawoud2021bringing, aafer2018precise} were proposed to infer the mappings between permissions and Android APIs via performing static analysis on Android framework. 
Felt et al.~\cite{felt2011android} proposed a technique {\mycode Stowaway} to construct permission specification, which used test generation to exercise Android APIs, and modified the internal permission check mechanism to log all permission checks on Android 2.2.
However, {\mycode Stowaway} only analyzed Android 2.2. 
Their obtained permission specification was precise but incomplete.
\textsc{PScout}~\cite{au2012pscout} produced a large permission specification for Android 4.1.1\textendash5.1.1, but its results are imprecise because its static analysis is coarse-grained.
\textsc{Copes}~\cite{bartel2012automatically} was proposed by Bartel et al. to discover over-privileged apps.
It also produced a permission specification using reachability analysis on Android framework.
Later in~\cite{bartel2014static}, Bartel et al. proposed an advanced class-hierarchy and field-sensitive analysis to extract the mappings between Android APIs and their required permissions.
They conducted the experiments on a large set of apps to detect \textit{permission gap} issues (for Android 2.2), i.e., there are more permissions than necessary.
Nguyen et al.~\cite{nguyen2017android} presented an approach to identifying Android APIs that access sensitive system or user data.
They analyzed six versions of Android (4.1.1\textendash6.0.1) and detected the APIs that may be sufficiently protected by permission checks.
\textsc{Axplorer}~\cite{backes2016demystifying} was built on top of \textsc{PScout} through more precise modeling of Android internals.
Yet, it contained few mappings for dangerous permissions.
Besides, Bao et al.~\cite{bao2017automated} proposed two approaches to recommending permissions required by APIs used in Android apps, via data mining and collaborative filtering techniques.
Compared with the above work, our work has a wider focus.
In addition to issues caused by the evolution of APIs and dangerous permissions, we also study many other types of ARP issues. 

\textbf{Evolution of permission mechanism.}
There are also a few studies that investigate Android's permission mechanism and its evolution.
Wei et al.~\cite{wei2012permission} studied a set of Android releases to investigate the evolution of the Android ecosystem to understand whether the permission model is allowing the platform and third-party apps to become more secure.
Alepis et al.~\cite{alepis2017hey} illustrated several drawbacks of old permission model (prior to Android 6.0) and highlighted the new features of the runtime permission model.
Zhauniarovich et al.~\cite{zhauniarovich2016small} conducted a comprehensive study to explore 
the changes of permission mechanism from Android 1.6 to Android 6.0.
Calciati et al.~\cite{calciati2017apps} conducted a preliminary study on over 14K releases of 227 Android apps, to understand how the apps evolved their permission requests across different releases, after migrating to the runtime permission model.
While these studies investigated the evolution of the permission mechanism, they have not systematically characterized the issues caused by such evolution at the code level.
In contrast, our work not only analyzed the evolution of permission mechanism from Android 6.0 to Android 10.0, but also investigated various types of ARP issues and built by far the most comprehensive issue taxonomy.


\vspace{-1mm}
\section{Conclusion and Future work}
In this work, we conducted a large-scale empirical study of ARP issues in the Android ecosystem.
We investigated 135 posts on Stack Overflow and 199 real ARP issues from 415 Android projects on GitHub.
Via analyzing these data, we built a taxonomy of ARP issues and studied their root causes, pervasiveness, seriousness, and fixing strategies.
In addition, we communicated with Android practitioners via online questionnaires and interviews to understand their common practices and experiences in terms of handing runtime permissions. 
From the studies, we obtained several important findings that can facilitate future research on Android runtime permissions and provide useful guidance to practitioners.
Futhermore, to understand the state of the art and the state of the practice, we investigated three existing ARP issue detectors.
With a carefully prepared benchmark, we experimentally evaluated these detectors and presented a detailed analysis of their limitations.
In future, we plan to address the limitations of existing detectors and design more effective techniques to help developers detect and fix ARP issues in their Android apps.





%

\ifCLASSOPTIONcaptionsoff
  \newpage
\fi



%

\bibliography{bibliography}{}
\bibliographystyle{IEEEtran}


%

\end{document}